\documentclass{iacrtrans}
\usepackage{cite}
\usepackage{multirow}
\usepackage{algorithm}
\usepackage{array}
\usepackage{booktabs}
\usepackage{amssymb}
\usepackage{color}
\usepackage{graphics}
\usepackage{adjustbox}
\usepackage{longtable}
\usepackage{hhline}
\usepackage{subcaption}
\usepackage{hyperref}
\usepackage{sectsty}
\usepackage[title]{appendix}
\usepackage[justification=raggedright,labelfont=bf,singlelinecheck=false]{caption}

\author{Malik Imran \and Zain Ul Abideen \and Samuel Pagliarini}
\institute{Centre for Hardware Security (CHS), Tallinn University of Technology (TalTech), Estonia, \email{{malik.imran,zain.abideen,samuel.pagliarini}@taltech.ee}}
\title[A Systematic Study of Lattice-based NIST PQC Algorithms: from Reference Implementations to Hardware Accelerators]{A Systematic Study of Lattice-based NIST PQC Algorithms: from Reference Implementations to Hardware Accelerators}

\begin{document}
\sloppy
\maketitle

\keywords[Post-quantum cryptography, NIST PQC algorithms, crypto-hardware, ASIC crypto accelerators]{Post-quantum cryptography \and NIST PQC algorithms
  \and crypto-hardware \and ASIC crypto accelerators}

\begin{abstract}
  Security of currently deployed public key cryptography algorithms is foreseen to be vulnerable against quantum computer attacks. Hence, a community effort exists to develop post-quantum cryptography (PQC) algorithms, i.e., algorithms that are resistant to quantum attacks. In this work, we have investigated how lattice-based candidate algorithms from the NIST PQC standardization competition fare when conceived as hardware accelerators. To achieve this, we have assessed the reference implementations of selected algorithms with the goal of identifying what are their basic building blocks. We assume the hardware accelerators will be implemented in application specific integrated circuit (ASIC) and the targeted technology in our experiments is a commercial 65nm node. In order to estimate the characteristics of each algorithm, we have assessed their memory requirements, use of multipliers, and how each algorithm employs hashing functions. Furthermore, for these building blocks, we have collected area and power figures for 12 candidate algorithms. For memories, we make use of a commercial memory compiler. For logic, we make use of a standard cell library. In order to compare the candidate algorithms fairly, we select a reference frequency of operation of 500MHz. Our results reveal that our area and power numbers are comparable to the state of the art, despite targeting a higher frequency of operation and a higher security level in our experiments. The comprehensive investigation of lattice-based NIST PQC algorithms performed in this paper can be used for guiding ASIC designers when selecting an appropriate algorithm while respecting requirements and design constraints.
\end{abstract}

\section{Introduction} \label{sec:Intro}
This is an era of internet-of-things (IoT), where all electronic devices around us, small and big, are connected to the internet, and hence, are vulnerable to an array of security threats. The backbone technology ensuring that sensitive data can be transmitted over an unsecured public channel is cryptography. Generally, it has two distinct flavours, i.e., private key and public key cryptography. Over the last few decades, public key cryptography (PKC) has become a fundamental security protocol for all forms of digital communication, both wired and wireless. It involves three main cryptographic operations, namely (1) public key-encryption/encapsulation (PKE), (2) key-exchange, and (3) digital signatures (DSA) \cite{Chen2016}. When performing these cryptographic operations, the security strength of currently deployed public key algorithms (e.g., RSA or Elliptic Curve Cryptography) is based on the difficulty of solving integer factorization and discrete logarithm problems. However, it has been showed that quantum computers can factorize integers in polynomial time -- the consequence being that traditional PKC algorithms may become vulnerable \cite{Shor1999}. Thus, to keep security of current communication practices, crypto researchers and developers are investigating different cryptographic “hard problems” (e.g., isogeny, lattices, multivariates etc.), in order to develop new algorithms that are robust against quantum computers.

Towards assessing different cryptographic methods against quantum attacks, the ongoing NIST post-quantum cryptography (PQC) standardization process serves as a beacon for the security community. In this text, we refer to algorithms participating in the process as \textbf{candidates} in a contest. Each candidate in the ongoing NIST contest implements one of two functions: (1) digital signature or (2) key encapsulation mechanism (KEM). At the onset of the contest, 82 algorithms were submitted for the standardization process. Based on the minimum acceptability criteria defined by NIST, 69 algorithms were considered for the first round. Considering several parameters (i.e., security, cost, performance, implementation characteristics, etc.), 43 and 11 algorithms were excluded at the completion of first and second rounds, respectively, while the remaining 15 algorithms were kept for third round \cite{Alagic2019}. 

The algorithms that remained in the second round can be categorized into five different cryptographic hard problems: (a) isogeny-based (1 algorithm), (b) lattice-based (12 algorithms), (c) code-based (7 algorithms), (d) multivariate polynomial cryptography (4 algorithms) and (e) hash-based digital signatures (2 algorithms) \cite{Alagic2019, CFD2011}. The security hardness of lattice-based cryptographic algorithms depends on solving the shortest vector problem (SVP), i.e., breaking an algorithm is the equivalent of solving the shortest vector problem \cite{SVP_Problem_in_Lattices}. Generally, the lattices are objects in an n-dimensional Euclidean space characterized by a regular arrangement of points, for complete mathematical formulations and constructions, interested readers can consult \cite{SVP_Problem_in_Lattices, Physical_Protection, LWR_Revisited}. 

A number of mathematical problems can be used to construct lattice-based schemes. However, the most commonly used mathematical problems are learning with errors (LWE) and learning with rounding (LWR). The LWE scheme is based on finding a vector $\boldsymbol{s}$ when given a matrix $\boldsymbol{A}$ and a vector $\boldsymbol{b=As+e}$, where $\boldsymbol{e}$ is a small error vector \cite{Physical_Protection}. On the other hand, the LWR problem is a variant of learning with errors where one replaces random errors with deterministic rounding \cite{LWR_Revisited}.

The following algorithms rely on the LWE problem: NTRU-Prime \cite{CFD2013}, FrodoKEM \cite{CFD2012}, NewHope \cite{CFD2015}, Crystals-KYBER \cite{CFD2018}, ThreeBears \cite{CFD2017}, LAC \cite{CFD2020}, NTRU \cite{CFD2023}, qTesla \cite{CFD2024} and Falcon \cite{CFD2026}. The LWR problem, on the other hand, is considered in Round5 \cite{CFD2021} and Saber \cite{CFD2022} algorithms. Finally, another popular mathematical problem to construct lattice-based scheme includes short vectors in lattices, as used in the Crystals-Dilithium \cite{CFD2025} algorithm. The aforementioned twelve algorithms were part of the NIST PQC standardization process and are the objects of this study.

Security is the primary evaluation criterion driving the NIST PQC competition and -- understandably -- the software implementations of the candidates focus on it. However, once the standardization process is over, the selected candidate(s) will surely be considered for hardware acceleration. In this scenario, a circuit designer will be tasked to improve performance, reduce area footprint, and reduce power consumption. Being so, it is imperative that we understand the constraints and characteristics of the algorithms when implemented as application specific integrated circuits (ASICs) such that we can then judge their feasibility of implementation for resource-constrained application domains. This is the main ambition of this work. It is important to highlight that 15 algorithms were selected (as finalists and alternate candidates) in the ongoing round 3 of the NIST PQC competition. Out of these 15 selected algorithms, 7 of them are lattice-based. Consequently,  we focus on lattice-based candidates as these are numerous and allow for richer comparisons.

\subsection{Existing studies and identified challenges} \label{sec:Intro_challenges}
Generally speaking, the last few years have brought a tremendous increase in the literature covering performance evaluations/comparisons of lattice-based cryptography. The algorithms considered in NIST's standardization process have received a fair share of the attention and have been implemented in different platforms \cite{Energy_Consumption_S/w,Pwr_Analysis_NTRU_Prime_S/w,H/w_study_of_sig_schemes_FPGA,Impl_Benchmark_FPGA,qTesla_FPGA,Architecture_of_NewHope_on_FPGA,Seven_lattice_Algos_FPGA_ASIC,PASS_FPGA_ASIC,SAPPHIRE,VPQC} (i.e., purely software, SW/HW co-design, FPGA, and ASIC). However, there are several shortcomings to these studies, even before we consider the specific challenges of ASIC implementation.

\textbf{Software-based implementations \cite{Energy_Consumption_S/w,Pwr_Analysis_NTRU_Prime_S/w}.} An evaluation study that focuses on the energy efficiency of software implementations is provided in \cite{Energy_Consumption_S/w}. A power analysis on software implementations of NTRU-Prime algorithm is presented in \cite{Pwr_Analysis_NTRU_Prime_S/w}. However, it must be emphasized that drawing power or energy figures from software implementations is not necessarily representative of what an ASIC implementation would eventually consume.

\textbf{FPGA-based implementations \cite{H/w_study_of_sig_schemes_FPGA,Impl_Benchmark_FPGA, qTesla_FPGA, Architecture_of_NewHope_on_FPGA}.} Leveraging high level synthesis (HLS), Verilog/VHDL codes are generated by the authors of \cite{H/w_study_of_sig_schemes_FPGA,Impl_Benchmark_FPGA} through which they evaluate different design characteristics (i.e., area, clock frequency, and number of cycles required for the overall computation). FPGA-specific Verilog code  were written by the authors of \cite{qTesla_FPGA, Architecture_of_NewHope_on_FPGA}. The Xilinx Artix-7 FPGA has often been used as a benchmarking platform for FPGA-based implementations. Therefore, in \cite{H/w_study_of_sig_schemes_FPGA}, qTesla and Crystals-Dilithium are evaluated on an Artix-7 device for key-pair generation, signature generation, and signature verification. An implementation and comparison study of few of the lattice-based algorithms is provided in \cite{Impl_Benchmark_FPGA}, where a Zynq UltraScale system-on-chip (SoC) platform has been utilized. An FPGA implementation of qTesla hash algorithm on Artix-7 is discussed in \cite{qTesla_FPGA}. In \cite{Architecture_of_NewHope_on_FPGA}, an efficient architecture for  NewHope is presented using a low-complexity Number Theoretic Transformation (NTT)/inverse NTT based modular multiplications.

\textbf{ASIC-based implementations \cite{Seven_lattice_Algos_FPGA_ASIC,PASS_FPGA_ASIC,SAPPHIRE,VPQC}.} Similar to FPGA-based implementations, Verilog/VHDL codes are generated through HLS by the authors of \cite{Seven_lattice_Algos_FPGA_ASIC,PASS_FPGA_ASIC} where they evaluate different design characteristics (i.e., area, clock frequency, and number of cycles required for the overall computation). Instead of describing dedicated cryptocores, solutions that make use of a RISC-V microprocessor are described in \cite{SAPPHIRE, VPQC}. Still referring to \cite{Seven_lattice_Algos_FPGA_ASIC}, while ASIC-specific discussion and results over 65nm standard cell library are reported for seven lattice-based algorithms (Saber, Crystals-KYBER, NewHope, FrodoKEM, NTRU, Crystals-Dilithium, and qTesla), there is no discussion on memories or multipliers (most likely abstracted by the use of HLS). We believe these are essential characteristics to be assessed and have carefully discussed both at length in this study. A design-space exploration of key generation, signature generation, and signature verification components of two digital signature algorithms (qTesla and Crystals-Dilithium) is presented in \cite{PASS_FPGA_ASIC} where the authors also make use of a 65nm standard cell library in their results. However, from the analysis of only two algorithms, it is hard to draw conclusions on their relative performance when compared to the many others. 

A configurable crypto-processor for post-quantum lattice-based protocols referred to as Sapphire has been presented in \cite{SAPPHIRE}, where the authors developed a dedicated instruction set, an arithmetic logical unit (ALU), and a control unit that interface with data and instruction memory. Moreover, the Sapphire processor was integrated with a RISC-V microprocessor to demonstrate FrodoKEM, NewHope, qTesla, Crystals-Kyber and Crystals-Dilithium algorithms. Similarly, an integrated domain-specific vector co-processor for post quantum cryptography algorithms with RISC-V microprocessor has been presented in \cite{VPQC}.

In summary, there is a real need for a study that systematically compares a large array of lattice-based PQC algorithms in the \textbf{fairest manner possible}. Our approach to address this problem is to compare all algorithms that were part of the second round of the NIST PQC competition and to perform this analysis by breaking down the algorithms into basic building blocks: memories, multipliers, and hashing cores. We opt not to make use of HLS in our study since HLS still is more convenient for FPGA-based implementations where BRAMs can be easily inferred. In ASICs, this method still presents some enormous challenges as the tools have no direct interface to proprietary memory compilers. This being said, we present our contributions in the next subsection.

\subsection{Our contributions} \label{sec:Intro_contributions}
The key contribution of this work is to provide a systematic study to investigate the required hardware resources for implementing PQC algorithms as ASIC hardware accelerators. We virtually conducted a competition within the NIST competition, albeit focusing only on lattice-based candidates.

Each algorithm submitted to the NIST PQC competition contains a set of building blocks and functions (e.g., arithmetic operators, logical operators, memory instances, transformations from one domain representation to another, etc.). These building blocks are not standardized as authors have the freedom to implement their C/C++ reference models in different ways. Therefore, we have defined a set of rules to fairly assess each reference implementation. By doing so, we define the essential building blocks of each algorithm in a consistent and fair fashion. Based on the defined principles, we have evaluated the reference implementations of the studied algorithms to (1) estimate the required memory sizes, (2) identify large arithmetic operators (i.e., multipliers) and (3) to identify the utilized hashing functions.

We proceed to compare the algorithms by making an ASIC-specific assessment using proprietary memory compilers and commercial standard cell libraries. The estimated area and power for ROMs and RAMs are calculated using memory compilers provided by a partner foundry. The target technology is 65nm bulk CMOS with a “low-power”  flavor. Amongst the identified arithmetic operators, we focus on the multipliers as they are often large and a bottleneck for the performance. Therefore, to calculate the actual hardware costs, we have developed Verilog RTL models for the multipliers and synthesized several variants of them using a commercial standard cell library. Furthermore, we have shown performance trends for different multiplier architectures over different input operand lengths ($2^1$ to $2^{12}$) in terms of power, area, and clock frequency. From these results, we then elect a single architecture for fair comparison among candidate algorithms.

Finally, we compile results for the combined memory and logic footprints of each studied algorithm. We provide the same result for power by utilizing a target frequency of 500MHz for all candidates. This is the information that we believe is the most useful for ASIC designers when selecting a PQC candidate to implement.
 
Our paper is organized as follows: a research protocol is defined in Section \ref{sec:Principles_def} to assess the reference implementations of selected algorithms. The characteristics of each assessed algorithm are described in Section \ref{sec:Assessment_of_Reference_Implementations}, where we identify memory instances, arithmetic operators (and the size of their operands), and hashing functions. Using memory compilers and standard cell libraries, required hardware resources are provided in Section \ref{sec:4}. The final evaluation of each studied algorithm, in terms of area and power, is presented in Section \ref{sec:5}. Finally, our conclusions are given in Section \ref{sec:conclusion}.

\section{Principles Definition} \label{sec:Principles_def}
We have defined a set of rules to select and evaluate the reference implementations of algorithms submitted to the NIST for PQC standardization process. These rules include the inclusion-exclusion principles (Section \ref{sec:InEx_principles}), selection of the algorithms for evaluations in this work (Section \ref{sec:Selection_algo}), criterion to estimate memory instances (Section \ref{sec:Memory_est}), and finally, the rules for estimating inputs and outputs of utilized arithmetic operators (Section \ref{sec:Rules_calc}).
\subsection{Inclusion-exclusion principles} \label{sec:InEx_principles}
We have defined the following principles for the inclusion-exclusion of a particular PQC algorithm in our study: 
\begin{itemize}
    \item \textbf{Participation in the NIST competition.} Include only algorithms that were considered on the NIST competition for standardization.
    \item \textbf{Underlying cryptographic primitive.} Include only algorithms that are built on the security problems of lattice-based cryptography.
    \item \textbf{Security levels.} For each studied algorithm, we consider only the reference model with  the highest security level.
    \item \textbf{Purpose of the algorithm.} For each particular algorithm, there might be a number of reference models that are developed either for encryption/decryption or key encapsulations/establishments. We opt not to include all the reference models as they serve inherently different purposes. Instead, we consider only the encryption/decryption reference models.
\end{itemize}

\subsection{Selection of algorithms and reference models} \label{sec:Selection_algo}
\begin{figure}[ht]
\centering \footnotesize
\includegraphics[width=4.8in,height=3.1in]{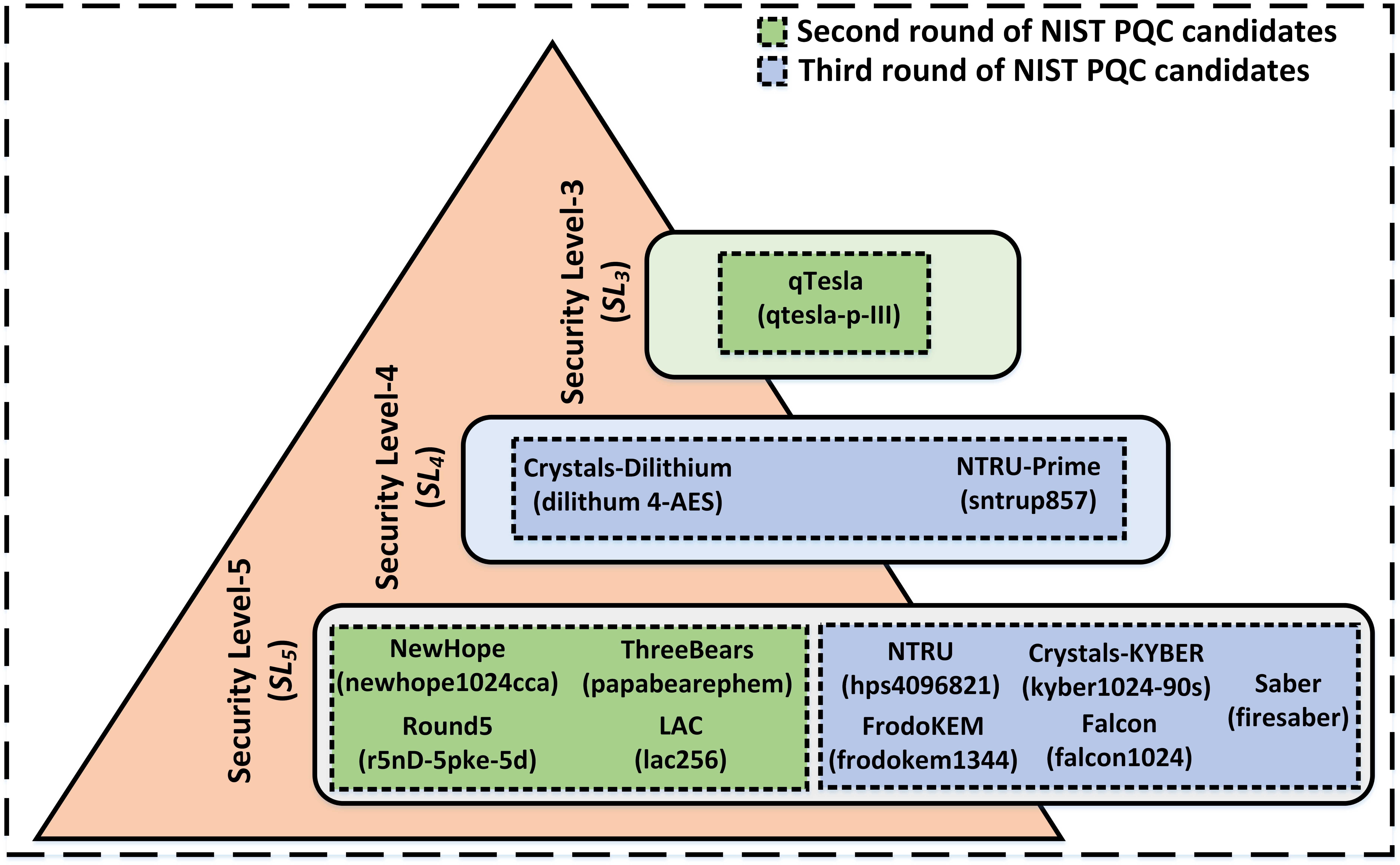}
\caption{Selected algorithms and the corresponding reference models utilized in this study.}
\centering
\label{fig:figure1}
\end{figure}

Based on the inclusion-exclusion principles defined above, we have selected twelve algorithms for this study. The details of these algorithms are given in Table \ref{tab:table1}. Note that the selected  algorithms have many different reference models targeted for different security levels. We denote these security levels as SL\textsubscript{i}. NIST has defined five different security levels for the standardization process: security levels SL\textsubscript{1}, SL\textsubscript{3}, and SL\textsubscript{5} are equivalent to security levels of AES-128, AES-192, and AES-256 bit key search. SL\textsubscript{2} and SL\textsubscript{4} are equivalent to SHA-256/SHA3-256 and SHA-384/SHA3-384 bit collision search. Consequently, based on the inclusion-exclusion principles, the selected reference models along with security levels for our evaluations are shown in Fig \ref{fig:figure1}.

\begin{table}[!]
\begingroup
\setlength{\tabcolsep}{1.5pt} 
\renewcommand{\arraystretch}{1.5} 
\fontsize{8}{10}\selectfont

\begin{longtable}{|p{2.2cm}|p{1.6cm}|p{3.0cm}|p{6.0cm}|}
\caption{Classification of reference models according to security levels}
\label{tab:table1} \\
\hline 
\parbox{2cm}{\textbf{Security Level (\textit{SL\textsubscript{i}})}} & 
\parbox{1.6cm}{\textbf{Type of\\ Algorithm}} & 
\parbox{3.0cm}{\textbf{Algorithm name}} & 
\parbox{6.0cm}{\textbf{Reference models}} \\ \hline

\multirow{1}{*}{\parbox{2.1cm}{\textbf{Security Level-0 (\textit{SL\textsubscript{0}})}}}  & PKE & Round5 & r5nD-0kem-2iot \vspace{2.3pt} \\ \hline
\multirow{10}{*}{\parbox{3.5cm}{\textbf{Security \\Level-1 (\textit{SL\textsubscript{1}})}}} &  & \parbox{3.0cm}{FrodoKEM} & frodokem640 \\ 
\cline{3-4} & & NewHope & \parbox{5.3cm}{newhope512cca, newhope512cpa } \\  
\cline{3-4} & & Crystals-KYBER & \parbox{5.5cm}{kyber512, kyber512-90s} \\  
\cline{3-4} & PKE & LAC & \parbox{5.3cm}{lac128} \\ 
\cline{3-4} & & Round5  & \parbox{6.0cm}{r5n1-1kem-0d, r5nD-1kem-0d, r5nD-1kem-4longkey, 5nD-1kem-5d, r5n1-1pke-0d, r5nD-1pke-0d, r5nD-5pke-5d} \\
\cline{3-4} & & Saber & \parbox{6cm}{Lightsaber} \\ 
\cline{3-4} & & NTRU & \parbox{6cm}{hps2048509} \\ 
\cline{2-4}&  & qTesla & \parbox{6cm}{qtesla-I, qtesla-I-s, qtesla-p-I} \\ 
\cline{3-4} & DSA & Crystals-Dilithium & \parbox{3.5cm}{dilithium1, dilithium1-AES} \\
\cline{3-4} & & Falcon & \parbox{6cm}{falcon512} \\ \hline
\multirow{5}{*}{\parbox{3.5cm}{\textbf{Security \\Level-2 (\textit{SL\textsubscript{2}})}}} & \multirow{2}{*}{PKE} & NTRU-Prime & sntrup653, ntrulpr653  \\ 
\cline{3-4} & & ThreeBears & babybear, babybearephem  \\ 
\cline{2-4} & \multirow{3}{*}{DSA} & qTesla & qtesla-II, qtesla-II-s \\
\cline{3-4} & & Crystals-Dilithium & dilithium2, dilithium2-AES \\ 
\cline{3-4} & & Falcon & falcon768 \\ \hline
\multirow{10}{*}{\parbox{3.5cm}{\textbf{Security \\Level-3 (\textit{SL\textsubscript{3}})}}} & \multirow{8}{*}{PKE} & NTRU-Prime & sntrup761, ntrulpr761  \\ 
\cline{3-4} & & FrodoKEM & frodokem976  \\ 
\cline{3-4} & & Crystals-KYBER & kyber768, kyber768-90s \\
\cline{3-4} & & LAC & lac192 \\
\cline{3-4} & & Round5 & \parbox{6cm}{r5n1-3kem-0d, r5nD-3kem-0d, r5nD-3kem-5d, r5n1-3pke-0d, r5n1-3pke-0smallCT, r5nD-3pke-0d, r5nD-3pke-5d} \\
\cline{3-4} & & Saber & Saber \\
\cline{3-4} & & NTRU & hps2048677, hrss701 \\
\cline{2-4} & \multirow{3}{*}{DSA} & 	 qTesla & qtesla-III, qtesla-III-s, qtesla-p-III \\
\cline{3-4} & & Falcon & qtesla-III, falcon768 \\
\cline{3-4} & & Crystals-Dilithium & dilithium3, dilithium3-AES \\ \hline
\multirow{4}{*}{\parbox{3.5cm}{\textbf{Security \\Level-4 (SL\textsubscript{4})}}} & \multirow{2}{*}{PKE} & NTRU-Prime & sntrup857, ntrulpr857  \\ 
\cline{3-4} & & ThreeBears & Mamabear  \\ 
\cline{2-4} & \multirow{2}{*}{DSA} & Crystals-Dilithium &dilithium4, dilithium4-AES	 \\
\cline{3-4} & & Falcon & falcon1024 \\ \hline
\multirow{11}{*}{\parbox{3.5cm}{\textbf{Security \\Level-5 (\textit{SL\textsubscript{5}})}}} & \multirow{8}{*}{PKE} & FrodoKEM & frodokem1344  \\ 
\cline{3-4} & & NewHope & newhope1024cca, newhope1024cpa  \\ 
\cline{3-4} & & Crystals-KYBER & kyber1024, kyber1024-90s  \\
\cline{3-4} & & ThreeBears & mamabearephem, papabear, papabearephem  \\
\cline{3-4} & & LAC & lac256  \\
\cline{3-4} & & Round5 & \parbox{6cm}{r5n1-5kem-0d, r5nD-5kem-0d, r5nD-5kem-5d, r5n1-5pke-0d, r5nD-5pke-0d, r5nD-5pke-5d}  \\
\cline{3-4} & & Saber & Firesaber  \\ 
\cline{3-4} & & NTRU & hps4096821  \\ 
\cline{2-4} & \multirow{2}{*}{DSA} & Crystals-Dilithium & dilithium4, dilithium4-AES	 \\
\cline{3-4} & & Falcon & falcon1024 \\ \hline
\end{longtable}
\endgroup
\end{table}

\subsection{Memory estimation criterion} \label{sec:Memory_est}
It is not uncommon for memories to take most of the area of a chip, especially in CPU-centric designs. Therefore, correctly estimating the number, type, and sizes of utilized memories is critical for assessing the size of an ASIC design. Memories are typically “placed by hand” and bear large influence on the floorplan of a chip. Therefore, it is important to correctly estimate the required memory instances using actual memory compilers. \\
In this work, we estimate the memories that would be required by all the selected reference models. We make a distinction between RAMs and ROMs because the compilers tend to be very different -- RAMs are typically an array of 6T/8T bitcells, while ROMs are typically via-programmable structures. In the studied reference models, variables to which only read operations are allowed are considered as a candidate for a ROM. Variables that present both read and write operations are considered as RAM candidates. However, not all variables are of interest for this exercise. For instance, variables that serve as flags or for temporary storage would not require a RAM as these would most likely reside in flip-flops or register banks. The same is true for small constants that do not require a ROM. Instead, these can be hardwired with tie cells.

\begin{figure}[ht]
\centering \footnotesize
\includegraphics[width=3in]{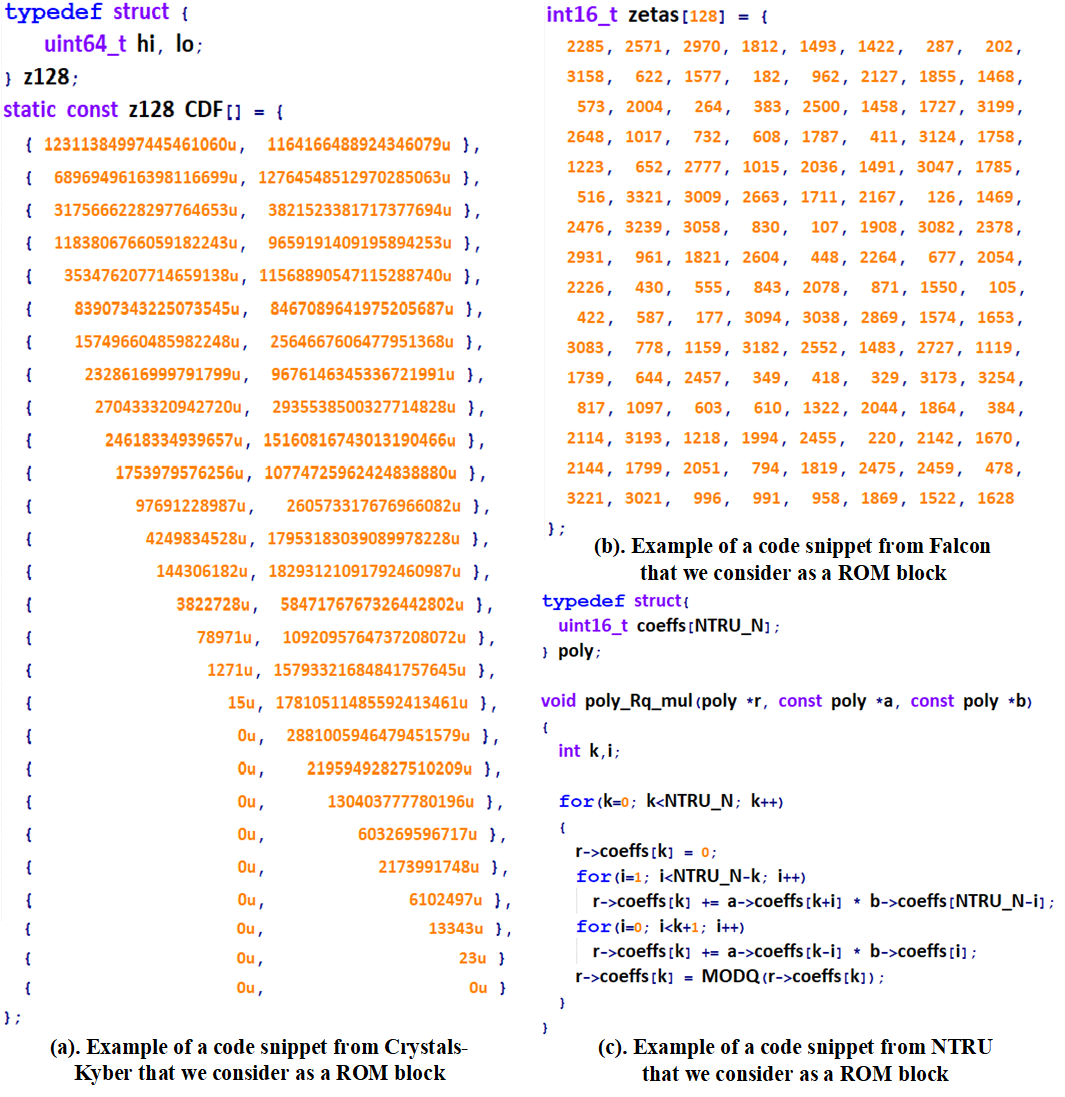}
\caption{Examples of code snippets from PQC candidates that explain how memory requirements were estimated.}
\centering
\label{fig:figure2}
\end{figure}

Moreover, different methods have been used to declare and initialize the memory instances in the analysed reference models. The most commonly used methods are (1) by scripting specific function, (2) by using arrays and (3) by introducing structures. This analysis becomes more complicated when dynamic memory allocation is employed. Different scenarios are highlighted in the code snippets shown in Fig. \ref{fig:figure2}.

As shown in Fig \ref{fig:figure2}(c), variables i and k are declared with the integer data type and therefore would be considered as flip-flops. It is important to mention one more time that we are interested in identifying actual memory instances, thus we ignore flip-flops at this step. For instance, a member of the structure \textit{poly} named \textit{coeffs} (see Fig \ref{fig:figure2}c) is considered as a RAM instance. To estimate the size of each variable of interest, we assess the total number of memory addresses indexed by it, as well as the number of bits stored at each address. Consequently, we have adopted the following methods to estimate the number of memory addresses in each memory instance and the number of stored bits at each memory address:

\textbf{Total number of memory addresses (\textit{p}).} In the reference implementations of selected reference models, both static and dynamic memory allocation methods have been used. For static allocation, memory size is known at compile time and observing the array dimension is sufficient, whereas for dynamic allocation we have to locate all malloc calls (direct or indirect). In Fig \ref{fig:figure2}(b), a ROM memory instance with 128 memory addresses (8 elements in each row $\times$ total number of rows, i.e., 16) is presented. Some memory instances are declared as structs and their size therefore depends on the members of the structure. The individual members of the structure, i.e., arrays, strings, etc., can each be of different types and lengths. For example, RAM memory instance in Fig \ref{fig:figure2}(c), a structure with name poly has one array type of member, i.e., \textit{coeffs [NTRU\textunderscore{N}]} where \textit{NTRU\textunderscore{N}} determines the length of array and its value defines the total number of memory addresses. Similarly, ROM memory instance in Fig \ref{fig:figure2}(a), a structure with name z128 have two array type of members, i.e., \textit{hi}, and \textit{lo}. Therefore, to evaluate structure type of memory instances we have considered each member of structure as an individual memory instance.

\textbf{Number of bits at each address (\textit{q}).} The utilized datatype determines the number of bits to be stored at each memory address. For example, \textit{uint16\textunderscore{t}} is utilized in the snippets of Fig. \ref{fig:figure2}(b) and (c), therefore each memory address can store 16 bits. It is important to emphasize that sometimes the code of a reference implementation may declare a large structure as an array/matrix of \textit{uint64\textunderscore{t}}, even if not all elements require 64 bits (see Fig. \ref{fig:figure2}(a)). In our evaluations, for a given memory instance, we have considered the length of all memory addresses the same.

\subsection{Rules for the calculation of operand sizes} \label{sec:Rules_calc}
We have explored the reference models provided by the corresponding authors for the identification of the length of different operands utilized in specific functions. This analysis is particularly important for arithmetical operations that deal with large-size inputs/outputs. Therefore, the total number of inputs and outputs are identified based on the parameters passed to the function of interest, while the size of each operand is identified based on its datatype. In a few cases, we had to resort to contacting the authors for clarification on the operand sizes. For the many hashing operations present in the reference implementations, sizes are already standardized by NIST itself and the effort lies in identifying the correct hashing function employed.

Now that we have defined rules to assess the building blocks of PQC algorithms (memory estimations, identification of arithmetic operators and their operand sizes, hashing functions, etc.), we are ready to assess the selected NIST PQC candidates as ASIC accelerators.

\section{Assessment of Building Blocks of the Selected NIST PQC Candidates} \label{sec:Assessment_of_Reference_Implementations}

\subsection{Memory estimations according to the defined criteria} \label{sec:Assessment_candidates}
Based on the principles defined in Section \ref{sec:Memory_est}, we provide sizes for the ROM and RAM instances in Tables \ref{tab:table2} and \ref{tab:table3}, respectively. Furthermore, we have not provided entries in Table \ref{tab:table2} for those algorithms where no ROMs have been identified, meaning that most likely the algorithm does not rely on precomputed constants. The first column of Tables \ref{tab:table2} and \ref{tab:table3} lists the name of the studied algorithm and the selected reference model. Starting from the second column, the meaning of each column is: (2) required memory instances (\textit{n}), (3) number of memory addresses per instance (\textit{p}), (4) number of bits stored at each address (\textit{q}), (5) size of each memory instance (\textit{r = p $\times$ q} in Kbytes), (6) size of \textit{n} memory instances (\textit{s = n $\times$ r} in Kbytes) and finally, (7) the total size (\textit{Total\textsubscript{size}}) is the sum of size for \textit{n} memory instances, i.e., (\textit{Total\textsubscript{size} = $\sum$ (s)} in Kbytes).

\begin{table}[h]
\begingroup
\setlength{\tabcolsep}{1.5pt} 
\renewcommand{\arraystretch}{1.5} 
\fontsize{8}{10}\selectfont

\begin{longtable}{|p{5.0cm}|l|l|l|l|l|l|}
\caption{Estimated sizes of required ROM instances} \label{tab:table2} \\ 
\hline
\parbox{5.0cm}{\textbf{Algorithm (reference model)}} & \parbox{1.1cm}{\textbf{n}} & \parbox{1.1cm}{\textbf{p}} & \parbox{1.1cm}{\textbf{q}} & \parbox{1.1cm}{\textbf{r}} & \parbox{1.1cm}{\textbf{s}} & \parbox{1.3cm}{\vspace{1mm} \textbf{Total\textsubscript{size}} (Kbytes) \vspace{1mm}} \\ \hline








\multirow{1}{*}{Crystals-KYBER (kyber1024-90s)} & \multirow{1}{*}{2} & \multirow{1}{*}{128} & \multirow{1}{*}{16} & \multirow{1}{*}{0.256} & \multirow{1}{*}{0.512} & \multirow{1}{*}{0.512} \\ \hline

\multirow{1}{*}{NewHope (newhope1024cca)} & \multirow{1}{*}{4} & \multirow{1}{*}{1024} & \multirow{1}{*}{16} & \multirow{1}{*}{2.048} & \multirow{1}{*}{8.192} & \multirow{1}{*}{8.192} \\ \hline

\multirow{2}{*}{LAC (lac256)} & 2 & 512 & 16 & 1.024 & 2.048 & \multirow{2}{*}{22.528} \\ 
\cline{2-6} & 1 & 5120 & 32 & 20.480 & 20.480 &  \\ \hline

\multirow{3}{*}{qTesla (qtesla-p-III)} & 1 & 444 & 32 & 1.776 & 1.776 & \multirow{3}{*}{22.000} \\ 
\cline{2-6} & 1 & 224 & 64 & 1.792 & 1.792 &  \\
\cline{2-6} & 2 & 2048 & 36 & 9.216 & 18.432 &  \\ \hline

\multirow{11}{*}{Falcon (falcon1024)} & 1 & 540 & 64 & 4.320 & 4.320 & \multirow{11}{*}{12.160} \\ 
\cline{2-6} & 1 & 1080 & 16 & 2.160 & 2.160 &  \\
\cline{2-6} & 2 & 31 & 64 & 0.248 & 0.496 &  \\
\cline{2-6} & 2 & 27 & 64 & 0.216 & 0.432 &  \\
\cline{2-6} & 2 & 30 & 64 & 0.240 & 0.480 &  \\
\cline{2-6} & 2 & 1024 & 16 & 2.048 & 4.096 &  \\
\cline{2-6} & 2 & 32 & 16 & 0.512 & 1.024 &  \\
\cline{2-6} & 2 & 64 & 16 & 1.024 & 2.048 &  \\
\cline{2-6} & 2 & 1024 & 8 & 1.024 & 2.048 &  \\
\cline{2-6} & 2 & 256 & 8 & 0.256 & 0.512 &  \\
\cline{2-6} & 2 & 512 & 8 & 0.512 & 1.024 &  \\ \hline
\end{longtable}
\endgroup
\end{table}

\begin{table}[!]
\begingroup
\setlength{\tabcolsep}{1.5pt} 
\renewcommand{\arraystretch}{1.5} 
\fontsize{8}{10}\selectfont

\begin{longtable}{|p{5.0cm}|l|l|l|l|l|l|}
\caption{Estimated sizes of required RAM instances} \label{tab:table3} \\ 
\hline
\parbox{5.0cm}{\textbf{Algorithm (reference model)}} & \parbox{1.1cm}{\textbf{n}} & \parbox{1.1cm}{\textbf{p}} & \parbox{1.1cm}{\textbf{q}} & \parbox{1.1cm}{\textbf{r}} & \parbox{1.1cm}{\textbf{s}} & \parbox{1.3cm}{\vspace{1mm} \textbf{Total\textsubscript{size}} (Kbytes) \vspace{1mm}} \\ \hline


\multirow{2}{*}{NTRU-Prime (sntrup857)} & 1 & 256 & 8 & 0.256 & 0.256 & \multirow{2}{*}{0.448} \\ 
\cline{2-6} & 1 & 24 & 64 & 0.192 & 0.192 &  \\ \hline

\multirow{2}{*}{FrodoKEM (frodokem1344)} & 3 & 10752 & 16 & 21.504 & 64.512 & \multirow{2}{*}{65.152} \\ \cline{2-6} 
& 5 & 64 & 16 & 0.128 & 0.640 &  \\ \hline

\multirow{11}{*}{Saber (firesaber)} & 1 & 32 & 8 & 0.032 & 0.032 & \multirow{11}{*}{1.888} \\ 
\cline{2-6} & 2 & 32 & 16 & 0.064 & 0.128 &  \\
\cline{2-6} & 1 & 128 & 8 & 0.128 & 0.128 &  \\
\cline{2-6} & 1 & 128 & 16 & 0.256 & 0.256 &  \\
\cline{2-6} & 1 & 64 & 8 & 0.064 & 0.064 &  \\
\cline{2-6} & 1 & 64 & 16 & 0.128 & 0.128 &  \\
\cline{2-6} & 1 & 4 & 512 & 0.256 & 0.256 &  \\
\cline{2-6} & 1 & 4 & 1024 & 0.512 & 0.512 &  \\
\cline{2-6} & 1 & 4 & 256 & 0.128 & 0.128 &  \\
\cline{2-6} & 1 & 4 & 512 & 0.256 & 0.256 &  \\ \hline

\multirow{1}{*}{NTRU (hps4096821)} & \multirow{1}{*}{14} & \multirow{1}{*}{821} & \multirow{1}{*}{16} & \multirow{1}{*}{1.642} & \multirow{1}{*}{22.988} & \multirow{1}{*}{22.988} \\ \hline

\multirow{5}{*}{ThreeBears (papabearephem)} & 1 & 40 & 8 & 0.040 & 0.040 & \multirow{5}{*}{3.409} \\ 
\cline{2-6} & 1 & 1584 & 8 & 1.584 & 1.584 &  \\
\cline{2-6} & 1 & 1697 & 8 & 1.697 & 1.697 &  \\
\cline{2-6} & 1 & 24 & 8 & 0.024 & 0.024 &  \\
\cline{2-6} & 2 & 32 & 8 & 0.032 & 0.064 &  \\ \hline

\multirow{2}{*}{Round5 (r5nD-5pke-5d)} & 2 & 16 & 8 & 0.016 & 0.032 & \multirow{2}{*}{0.064} \\ 
\cline{2-6} & 1 & 32 & 8 & 0.032 & 0.032 &  \\ \hline

\multirow{1}{*}{Crystals-Dilithium (dilithium4-AES)} & \multirow{1}{*}{3} & \multirow{1}{*}{256} & \multirow{1}{*}{32} & \multirow{1}{*}{1.024} & \multirow{1}{*}{3.072} & \multirow{1}{*}{3.072} \\ \hline

\multirow{2}{*}{Crystals-KYBER (kyber1024-90s)} & 5 & 256 & 16 & 0.512 & 2.560 & \multirow{2}{*}{2.816} \\ 
\cline{2-6} & 1 & 128 & 16 & 0.256 & 0.256 &  \\ \hline

\multirow{1}{*}{NewHope (newhope1024cca)} & \multirow{1}{*}{8} & \multirow{1}{*}{1024} & \multirow{1}{*}{16} & \multirow{1}{*}{2.048} & \multirow{1}{*}{16.384} & \multirow{1}{*}{16.384} \\ \hline

\multirow{3}{*}{LAC (lac256)} & 1 & 2080 & 8 & 2.080 & 2.080 & \multirow{3}{*}{4.560} \\ 
\cline{2-6} & 1 & 1056 & 8 & 1.056 & 1.056 &  \\
\cline{2-6} & 1 & 1024 & 8 & 1.424 & 1.424 &  \\ \hline

\multirow{5}{*}{qTesla (qtesla-p-III)} & 1 & 2048 & 8 & 2.048 & 2.048 & \multirow{5}{*}{152.576} \\ 
\cline{2-6} & 1 & 9600 & 32 & 38.400 & 38.400 &  \\
\cline{2-6} & 1 & 10240 & 32 & 49.960 & 40.960 &  \\
\cline{2-6} & 1 & 1408 & 32 & 5.632 & 5.632 &  \\
\cline{2-6} & 4 & 2048 & 64 & 16.384 & 65.536 &  \\ \hline

\multirow{2}{*}{Falcon (falcon1024)} & 5 & 1024 & 16 & 2.048 & 10.240 & \multirow{2}{*}{22.744} \\ 
\cline{2-6} & 6 & 521 & 32 & 2.084 & 12.504 &  \\ \hline
\end{longtable}
\endgroup
\end{table}
In the text that follows, we describe where the memory requirements are coming from for each algorithm.\\
\begin{itemize}

 \item \textbf{NTRU-Prime.} The implementation of NTRU-Prime algorithm requires only 2 RAM instances. One RAM instance with a size of 0.256 Kbytes and another instance with a size of 0.192 Kbytes is required to store intermediate and final results of arithmetic (modular addition and subtraction) and logical operations that take place during the algorithm.
 
 \item \textbf{FrodoKEM.} The implementation of FrodoKEM requires 8 RAM instances. Three RAM instances with identical sizes of 21.504 Kbytes are required to perform modular addition and subtraction operations over matrices of size \textit{M $\times$ N}, where \textit{M} is the number of rows (1344) and \textit{N} is the number of columns (either 1344 or 8). The remaining (five) RAM instances with sizes of 0.128 Kbytes each are required to perform addition and subtraction operations of \textit{M $\times$ N} matrices, when value for both \textit{M} and \textit{N} equals 8.
 
 \item \textbf{Saber.} The Saber algorithm requires 11 RAM instances for its implementation. Two RAM instances with sizes of 0.064 Kbytes each are required to pack and unpack 3 bits. A RAM instance with a size of 0.032 Kbytes is required to implement the transformation function \textit{BS2POL()} of byte string into polynomial. Two RAM instances of sizes 0.128 Kbytes and 0.256 Kbytes are required to pack and unpack 4 bits. To pack and unpack 6 bits, two RAM instances of sizes 0.064 Kbytes and 0.128 Kbytes are required. Moreover, four RAM instances of sizes 0.256 Kbytes, 0.512 Kbytes, 0.128 Kbytes and 0.256 Kbytes are required to implement the transformation functions, i.e., \textit{POLVEC\textsubscript{N}2BS()} and \textit{BS2POLVEC\textsubscript{N}()}. The  \textit{POLVEC\textsubscript{N}2BS()} function takes a vector as an input and transforms it into a byte string while the function \textit{BS2POLVEC\textsubscript{N}()} takes a byte string as an input and transforms it into a vector.
 
 \item \textbf{NTRU.} For implementation of NTRU, 14 RAM instances are required. All the RAM instances with sizes of 1.642 Kbytes each are required to keep intermediate and final results of the NTRU algorithm.
 
 \item \textbf{ThreeBears.} The reference implementation of ThreeBears requires only 6 RAM instances. Two RAM instances with sizes of 0.040 Kbytes and 1.584 Kbytes are required to hold private and public keys, respectively. Two RAM instances with sizes of 1.697 Kbytes and 0.024 Kbytes are required to hold capsule and message seed bytes. The remaining two RAM instances with sizes of 0.032 Kbytes each are required to keep encapsulation seed bytes and shared secret bytes.
 
 \item \textbf{Round5.} The implementation of Round5 requires only 3 RAM instances. Two RAM instances of size 0.016 Kbytes each are required to provide inputs and get outputs to/from AES cores \cite{AES_CORE}. The remaining RAM instance with a size of 0.032 Kbytes is required to keep the key (outside the AES core) for the execution of AES algorithm. It is important to mention that the structure of Round5 algorithm is flexible and its security strength is based on defining the LWR problem \cite{CFD2021}.
 
 \item \textbf{Crystals-Dilithium.} The implementation of Crystals-Dilithium requires only 3 RAM instances to keep intermediate results for polynomial addition and subtraction operations. All three instances require 1.024 Kbytes.
 
 \item \textbf{Crystals-KYBER.} The reference model of Crystals-KYBER algorithm requires 2 ROM and 6 RAM instances, respectively. Two ROM instances of size 0.256 Kbytes each are required to keep original and inverse of original Zeta values, which are required for the NTT computations. Five RAM instances with sizes of 0.512 Kbytes each are required to decompress the polynomials (1 instance), to perform polynomial arithmetic operations -- modular addition and subtraction (3 instances), and to convert polynomial coefficients for modular multiplications from Montgomery to normal domain (1 instance). The remaining RAM instance with size 0.256 Kbytes is required to convert bytes to polynomials.
 
 \item \textbf{NewHope.} The reference implementation of NewHope algorithm requires 4 ROM and 8 RAM instances. The four ROM instances with sizes of 2.048 Kbytes each are required to keep pre-computed constant values for executing, (1) bitrev\textunderscore table, (2) omegas \textunderscore inv\textunderscore bitrev\textunderscore montgomery, (3) gammas\textunderscore bitrev\textunderscore montgomery and (4) gammas\textunderscore inv\textunderscore montgomery. The bitrev\textunderscore table contains 10-bit indices and is required to re-order the polynomials. The omegas\textunderscore inv\textunderscore bitrev\textunderscore montgomery contains inverses of powers of $n^{th}$ root of unity in Montgomery domain with R = $2^{18}$ in bit reversed order. The gammas\textunderscore bitrev\textunderscore montgomery keeps powers of $n^{th}$ root of -1 in Montgomery domain with R = $2^{18}$ in bit reversed order. The gammas \textunderscore inv\textunderscore montgomery holds inverses of powers of $n^{th}$ root of -1 divided by n in Montgomery domain with R = $2^{18}$. The RAM instances (total=8), each sized as 2.048 Kbytes, are required to compute arithmetic operations (instead of modular multiplications) over polynomial representations.
 
 \item \textbf{LAC.} The LAC algorithm requires 3 ROM and 3 RAM instances for its implementation. Two ROM instances with sizes of 1.024 Kbytes each are required to keep initial power and log values for the polynomial computations. One ROM instance with a size of 20.480 Kbytes is required to hold modulus values required for the reduction of polynomials after multiplication operation. Three RAM instances with sizes of 2.080 Kbytes, 1.056 Kbytes and 1.424 Kbytes are required to keep bytes of generated secret key, public key and the cipher texts.
 
 \item \textbf{qTesla.} The implementation of qTesla requires 4 ROM and 8 RAM instances. ROM instances with sizes of 1.776 Kbytes, 1.792 Kbytes, 9.216 Kbytes, and 9.216 Kbytes are required to hold initial values for the Gaussian sampler with 32-bit words, Gaussian sampler with 64-bit words, constants for the Zeta computations, and constants for the inverse Zeta computations, respectively. One RAM instance with a size of 2.048 Kbytes is required to pack the secret key. Two RAM instances with sizes of 38.400 Kbytes and 40.960 Kbytes are required to encode and decode the public key. Two instances of RAMs with sizes of 5.632 Kbytes and 16.384 Kbytes are required to encode and decode sampled message. Three RAM instances with sizes of 16.384 Kbytes each are required to perform polynomials addition and subtraction operations.
 
 \item \textbf{Falcon.} The reference implementation of Falcon requires 20 ROM and 13 RAM instances. Two ROM instances with sizes of 4.32 Kbytes and 2.16 Kbytes are required to keep initial values for discrete Gaussian distributions and bit reversal index table, respectively. Two ROM instances with sizes of 0.248 Kbytes each are required to hold precomputed continuous cumulative distribution function (CoDF) values. For small primes, two ROM instances with sizes of 0.216 Kbytes each are required to hold precomputed cumulative distribution function (CDF) values. Similarly, for large primes, two ROM instances with sizes of 0.240 Kbytes each are required to hold precomputed CDF values. For binary polynomials, two ROM instances with the same size of 2.048 Kbytes are required to keep initial values for the computation of NTT and inverse NTT operations. For ternary polynomials, four ROM instances with sizes of 0.512 Kbytes (2 instances) and 1.024 Kbytes (2 instances) are required to keep initial values for the computation of NTT and inverse NTT operations. For complex number computations, a total of six ROM instances with sizes of 1.024 Kbytes (2 instances), 0.256 Kbytes (2 instances) and 0.512 Kbytes (2 instances) are required to keep initial values of real and imaginary parts of normal, binary, and cubic/ternary polynomials. Five RAM instances, each sized 2.048 Kbytes, are required to generate initial parameters using Gaussian distribution function. Six RAM instances with sizes of 2.084 Kbytes each are required to generate small primes (namely \textbf{p}, \textbf{g} and \textbf{s} in the reference implementation) for both binary and ternary polynomials.
\end{itemize}

As summarized, the requirements for ROM and RAM sizes are relatively small by modern software standards. For instance, for the algorithm with the highest memory requirement, qTesla, the values are well under 1Mbyte. However, when considering ASIC implementations, these memory sizes are not modest and may render some algorithms severely less attractive than others. In order to give a better understanding of the magnitude of the memory sizes, we later provide area values in Tables \ref{tab:table6} and \ref{tab:table7} of Section \ref{sec:4_memcalc}.

\subsection{Common arithmetic operators and operand sizes} \label{sec:Arithmetic_sizes}
According to the rules defined in Section \ref{sec:Rules_calc}, the identified common arithmetic operators and the length of input and output operands are shown in Table \ref{tab:table4}. In the first column, the name of particular algorithm and its reference model is presented. The second column is further divided into three subcolumns: (1) identified common arithmetic operators, (2) used function name in the provided reference implementations and the name of algorithm/method used for implementations (in parentheses) and finally, (3) the length of required input (\textit{OP\textsubscript{1}}, \textit{OP\textsubscript{2}} \& \textit{OP\textsubscript{3}}) and output (\textit{OP\textsubscript{4}}) operands in bits. Our analysis focuses on arithmetic operators of large sizes and of high computational complexity. Naturally, multiplication stands out among those operations listed in Table \ref{tab:table4}.

\begin{table}[!]
\begingroup
\setlength{\tabcolsep}{1.5pt} 
\renewcommand{\arraystretch}{1.5} 
\fontsize{8}{10}\selectfont
\begin{longtable}{|p{3cm}|p{1.8cm}|p{4cm}|p{1cm}|p{1cm}|p{1cm}|p{1cm}|}
\caption{Required common arithmetic operators and the length of input and output operands} \label{tab:table4} \\
\hline 

\multirow{3}{*}{\parbox{3cm}{\textbf{Algorithm\\(reference model)}}} & \multicolumn{5}{l}{\textbf{Details of arithmetic operators and operand lengths (in bits)}} & \\ \cline{2-7}
& \multirow{2}{*}{\textbf{Operators}} & \multirow{2}{*}{\textbf{Function name (method)}} & \multicolumn{4}{l|}{\textbf{Input/Output operands}} \\ \cline{4-7}
& & & \textit{OP\textsubscript{1}} & \textit{OP\textsubscript{2}} & \textit{OP\textsubscript{3}} & \textit{OP\textsubscript{4}} \\ \hline

\multirow{2}{*}{\parbox{3cm}{NTRU-Prime (sntrup857)}} & \multirow{2}{*}{A $\times$ B}  & {\textit{Rq\textunderscore{mult}\textunderscore{small()}} (SBM)} & 12176 & 6088 & - & 12176 \\ \cline{3-7}
&  & \textit{Rq\textunderscore{mult()}} (SBM) & 6088 & 6088 & - & 6088  \\ \hline

\multirow{6}{*}{\parbox{3cm}{FrodoKEM (frodokem1344)}} & $(C^{-} \times A^{-}) + B^{-}$  & \textit{frodo\textunderscore{mul}\textunderscore{add}\textunderscore{as}\textunderscore{plus}\textunderscore{e}()} (SBM) & 172032 & 172032 & 128 & 172032 \\ \cline{2-7}
 & $(C^{-} \times A^{-}) + B^{-}$  & \textit{frodo\textunderscore{mul}\textunderscore{add} \textunderscore{sa}\textunderscore{plus}\textunderscore{e}()} (SBM) & 172032 & 172032 & 128 & 172032 \\ \cline{2-7} 
 & $(A^{-} \times B^{-}) + C^{-}$  & \textit{frodo\textunderscore{mul}\textunderscore{add} \textunderscore{sb}\textunderscore{plus}\textunderscore{e}()} (SBM) & 172032 & 172032 & 128 & 128 \\ \cline{2-7} 
 & $A^{-} \times B^{-}$  & \textit{frodo\textunderscore{mul} \textunderscore{bs}()} (SBM) & 172032 & 172032 & 128 & 128 \\ \cline{2-7} 
 & $A^{-} + B^{-}$ & \textit{poly\textunderscore{add}()} (SBM) & 128 & 128 & - & 128 \\ \cline{2-7} 
 & $A^{-} - B^{-}$  & \textit{poly\textunderscore{sub}()} (SBM) & 128 & 128 & - & 128 \\ \hline

\multirow{2}{*}{\parbox{3cm}{Saber (firesaber)}} & \multirow{2}{*}{$(A \times B)$} & \textit{karatsuba\textunderscore{simple()}} (KM) & \multirow{1}{*}{4096} & \multirow{1}{*}{4096} & \multirow{1}{*}{-} & \multirow{1}{*}{4096} \\ \cline{3-7}
& & \textit{toom\textunderscore{cook}\textunderscore{4way}()} (TCM) & 4096 & 4096 & - & 4096 \\ \hline

\multirow{3}{*}{\parbox{3cm}{NTRU\\(hps4096821)}} & \multirow{2}{*}{$(A \times B)$} & \textit{poly\textunderscore{Rq}\textunderscore{mul}()} (KM) & 11216 & 11216 & - & 11216  \\ \cline{3-7}
&  & \textit{poly\textunderscore{Sq}\textunderscore{mul}()} (TCM) & 11216 & 11216 & - & 11216  \\ \cline{2-7}
& $1/A$ & \textit{poly\textunderscore{S3}\textunderscore{inv}()} (Almost algo) & 11216 & - & - & 11216 \\ \hline

\multirow{2}{*}{\parbox{3cm}{ThreeBears\\(papabearephem)}} & \multirow{2}{*}{+= $A \times B$}  & \multirow{2}{*}{\textit{mac()} (2-Way KM)} & \multirow{2}{*}{3120} & \multirow{2}{*}{3120} & \multirow{2}{*}{-} & \multirow{2}{*}{3120} \\
& & & & & & \\ \hline

\multirow{2}{*}{\parbox{3cm}{Round5\\(r5nD-5pke-5d)}} & $(A^{-} \times B^{-})$ & \textit{ringmul\textunderscore{p()}} (SBM) & 15136 & 6208 & - & 7840 \\ \cline{2-7}
& $(A^{-} \times B^{-})$ & \textit{ringmul\textunderscore{p()}} (SBM) & 15136 & 6208 & - & 15136 \\ \hline

\multirow{3}{*}{\parbox{3cm}{Crystals-Dilithium (dilithium4-AES)}} & $(A + B)$ & \textit{poly\textunderscore{add}()} (SBM) & 8192 & 8192 & - & 8192  \\ \cline{2-7}
& $(A - B)$ & \textit{poly\textunderscore{sub}()} (SBM) & 8192 & 8192 & - & 8192  \\ \cline{2-7}
& $(A \times B)$ & \parbox{4cm}{\textit{poly\textunderscore{pointwise}
\textunderscore{invmontgomery}()} (NTT)} & 8192 & 8192 & - & 8192  \\ \hline

\multirow{3}{*}{\parbox{3cm}{Crystals-KYBER (kyber1024-90s)}} & $A \times B$  & \textit{poly\textunderscore{basemul}()} (NTT) & 3072 & 3072 & - & 3072 \\ \cline{2-7}
& $A^{-} + B^{-}$ & \textit{poly\textunderscore{add}()} (SBM) & 3072 & 3072 & - & 3072 \\ \cline {2-7}
& $A^{-} - B^{-}$ & \textit{poly\textunderscore{sub}()} (SBM) & 3072 & 3072 & - & 3072 \\ \hline

\multirow{3}{*}{\parbox{3cm}{NewHope (newhope1024cca)}} & $A \times B$  & \textit{poly\textunderscore{mul} \textunderscore{pointwise}()} (NTT) & 16384 & 16384 & - & 16384 \\ \cline{2-7} 
& $A^{-} + B^{-}$ & \textit{poly\textunderscore{add}()} (SBM) & 128 & 128 & - & 128 \\ \cline{2-7}
& $A^{-} - B^{-}$  & \textit{poly\textunderscore{sub}()} (SBM) & 128 & 128 & - & 128 \\ \hline 

\multirow{2}{*}{\parbox{3cm}{LAC (lac256)}} & $A \times B$  & \textit{poly\textunderscore{aff()}} (SBM) & 8192 & 8192 & 8192 & 8192 \\ \cline{2-7}
& $(A \times B)$ + C  & \textit{poly\textunderscore{mul()}} (SBM) & 8192 & 8192 & 32 & 8192  \\ \hline

\multirow{3}{*}{\parbox{3cm}{qTesla\\(qtesla-p-III)}} & $(A \times B)$ & \textit{poly\textunderscore{mul}()} (NTT) & 16384 & 16384 & - & 16384  \\ \cline{2-7}
& $(A + B)$ & \textit{poly\textunderscore{add}()} (SBM) & 16384 & 16384 & - & 16384  \\
\cline{2-7}
& $(A - B)$ & \textit{poly\textunderscore{sub}\textunderscore{reduce}()} (SBM) & 16384 & 16384 & - & 16384  \\ \hline

\multirow{3}{*}{\parbox{3cm}{Falcon (falcon1024)}} & $(A - B)$ & \textit{mq\textunderscore{poly}\textunderscore{sub}()} (SBM) & 24576 & 24576 & - & 24576  \\
\cline{2-7} 
& $(A \times B)$ & \textit{mq\textunderscore{poly}\textunderscore{montymul}\textunderscore{ntt}()} (Mont) & 24576 & 24576 & - & 24576  \\ \cline{2-7} 
& $(A \times B)$ & \textit{mq\textunderscore{montymul}()} (Mont) & 32 & 32 & 32 & 32  \\ \hline
\end{longtable}
\endgroup
\end{table}
As shown in the second column of Table \ref{tab:table4}, operations such as addition, subtraction, and multiplication are frequently used in the reference implementations. Apart from the identified operators in Table \ref{tab:table4}, there are also various other operators such as logical, arithmetic (inversion/division), reduction (modular and Montgomery), etc., that are used in the reference implementations to perform different functions. For instance, there are several functions that handle coordinate conversions/re-conversions from one domain to another (classical/affine to NTT or Montgomery and vice versa). Additionally, the NIST PQC candidates require different sampler functions for multiple purposes such as generating public parameters in a particular algorithm. The arithmetic additions and subtraction operations, shown in column three of Table \ref{tab:table4}, are commonly implemented by using schoolbook method.
 
It is worth mentioning that several algorithms make use of rather large operands (see fourth column of Table \ref{tab:table4}). This being said, it is the multiplication operation that becomes the most computationally intensive operation and this is reflected by the fact that several authors carefully optimised their codes for handling the multiplication of such large operand sizes. Several different methods are considered, including schoolbook \cite{Liu2019}, 2-Way Karatsuba \cite{Kashif2019}, 4-Way Toom-Cook \cite{Bodrato2007}, NTT \cite{10.1007/978-3-319-48965-0_8} and Montgomery \cite{10.1007/3-540-48059-5_9} methods. Consequently, in the text that follows, we discuss the identified multiplication methods utilized by the candidates of the NIST PQC competition.

\textbf{Schoolbook method (SBM).} The schoolbook method by generating partial products and matrix multiplications is often applied to lattice-based cryptography \cite{Liu2019}. In the studied algorithms, it was applied in \cite{CFD2013}, \cite{CFD2012}, \cite{CFD2020}, \cite{CFD2021} and \cite{CFD2023}. \textbf{Generating partial products.} Multiplications by generating partial products is implemented in NTRU-Prime, LAC, and NTRU algorithms. In NTRU-Prime, the identified multiplier functions, i.e., \textit{R3\textunderscore{mult()}} and \textit{Rq\textunderscore{mult}\textunderscore{small()}}, make use of SBM by generating partial products along with shifting and addition operations. For LAC algorithm, the \textit{poly\textunderscore{mul()}} and \textit{poly\textunderscore{aff()}} functions are implemented to compute polynomial multiplications using shifting and addition operations. The function \textit{poly\textunderscore{aff()}} computes multiplication over addition, i.e., $(A \times B)+ C$, while in \textit{poly\textunderscore{mul()}} function, only the polynomial multiplication is performed. The size of the operands in the LAC algorithm is 8192 bits. In both NTRU-Prime and LAC algorithms, a modular reduction mod\textunderscore{q} is applied to reduce the resultant polynomial. In NTRU algorithm, multiplication using shifting and addition operations along with mod\textunderscore{3} and mod\textunderscore{q} reductions are implemented in \textit{poly\textunderscore{Rq}\textunderscore{mul()}}, \textit{poly\textunderscore{Sq}\textunderscore{mul()}} and \textit{poly\textunderscore{S3}\textunderscore{mul()}} functions. Furthermore, an almost inverse algorithm is implemented in \textit{poly\textunderscore{S3}\textunderscore{inv()}} function for the computations of polynomial inversion operation. For both multiplication and inversion operations, inputs and outputs are 11216 bits in length. \textbf{Matrix multiplications.} Matrix multiplication using schoolbook method is considered in FrodoKEM and Round5 algorithms. For FrodoKEM algorithm, matrix multiplications are implemented in \textit{frodo\textunderscore{mul}\textunderscore{add}\textunderscore{as}\textunderscore{plus}\textunderscore{e()}}, \textit{frodo\textunderscore{mul}\textunderscore{bs()}}, \textit{frodo\textunderscore{mul}\textunderscore{add}\textunderscore{sa}\textunderscore{plus}\textunderscore{e()}}, and \textit{frodo\textunderscore{mul}\textunderscore{add}\textunderscore{sb}\textunderscore{plus}\textunderscore{e()}} functions. The row-wise multiplications are performed in \textit{frodo\textunderscore{mul}\textunderscore{add}\textunderscore{as}\textunderscore{plus}\textunderscore{e()}} and \textit{frodo\textunderscore{mul}\textunderscore{bs()}} functions while in \textit{frodo\textunderscore{mul}\textunderscore{add}\textunderscore{sa}\textunderscore{plus}\textunderscore{e()}} and \textit{frodo\textunderscore{mul}\textunderscore{add}\textunderscore{sb}\textunderscore{plus}\textunderscore{e()}} functions, column-wise multiplications are implemented. Furthermore, in the function \textit{$frodo\_mul\_add\_as\_plus\_e()$}, the second operand, i.e., \textit{OP\textsubscript{2}}, is multiplied with the first operand, i.e., \textit{OP\textsubscript{1}}, and then added to the third operand, i.e., \textit{OP\textsubscript{3}}, to generate the final output. In the function \textit{frodo\textunderscore{mul}\textunderscore{bs()}}, \textit{OP\textsubscript{2}} is multiplied with the first operand, i.e., \textit{OP\textsubscript{1}}, and the resultant product is stored. The function \textit{frodo\textunderscore{mul}\textunderscore{add}\textunderscore{sa}\textunderscore{plus}\textunderscore{e()}} allows multiplication of \textit{OP\textsubscript{1}} by \textit{OP\textsubscript{2}} and then adds \textit{OP\textsubscript{3}} to the product to generate the final output. The function \textit{frodo\textunderscore{mul}\textunderscore{add}\textunderscore{sb}\textunderscore{plus}\textunderscore{e()}} allows multiplication of \textit{OP\textsubscript{2}} by \textit{OP\textsubscript{1}}, and then adds \textit{OP\textsubscript{3}} to the product to generate the final output. In Round5 algorithm, row-wise matrix multiplication method is implemented in \textit{ringmul\textunderscore{p()}} and \textit{ringmul\textunderscore{q()}} functions. FrodoKEM uses 16 bit data types in reference implementations, so a modulo $2^{16}$ is used for modular reduction (see section 1.2.3 in reference \cite{CFD2012}) while in Round5, mod\textit{\textsubscript{q}} reduction is implemented (see section 2.2 in reference \cite{CFD2021}).  

\textbf{Karatsuba (KM) \cite{Kashif2019,article25} and Toom-Cook (TCM) \cite{Bodrato2007} multipliers.} Both KM and TCM are based on the idea of splitting input operands into \textit{n} parts, where \textit{n} determines the length of each split operand. Then, the multiplication of each split part (\textit{n}) is recursively computed using addition and shift operations. For mathematical formulations, interested readers can turn to \cite{Kashif2019,Bodrato2007,article25}. Functions named \textit{karatsuba\_simple()} and \textit{toom\_cook\_4way()} are implemented in Saber to perform multiplications using 2-Way KM and 4-Way TCM with mod\textit{\textsubscript{q}} reduction. In \cite{CFD2022}, the terms 2-Way and 4-Way imply that the input operands are split into 2 and 4 equal parts for multiplication and the required operands length for multiplication computations are 4096 bits. The multiply and accumulate function, i.e., \textit{mac()}, is implemented in ThreeBears for combined polynomial multiplication and addition using a 2-Way KM. It takes input operand sizes of 3120 bits and generates 3120 bits as an output after modulo reduction (it uses mod(n)).

\textbf{NTT method \cite{harvey_2013}.} Discrete Fourier transform (DFT) has numerous applications in signal processing and multiplication of large numbers or operands. NTT is the generalized form of DFT which exploits the convolution feature to multiply large operands \cite{10.1007/978-3-319-48965-0_8}. For comprehensive mathematical formulations and general understanding of transformations over NTT-based multiplication method, we refer to \cite{harvey_2013}. The NTT method for polynomial multiplications with Montgomery reduction is utilized in NewHope, Crystals-KYBER, qTesla, and Crystals-Dilithium algorithms. The function named \textit{poly\_mul\_pointwise()} is implemented in NewHope for the computation of multiplication and it takes operands that are 16384 bits long. In Crystals-KYBER, qTesla, and Crystals-Dilithium algorithms, the corresponding multiplication functions are named \textit{poly\_basemul()}, \textit{poly\_mul()} and \textit{poly\_pointwise\_invmontgomery()}, respectively. Furthermore, each polynomial in the Crystals-KYBER algorithm has 256 coefficients and each coefficient is in the set {0, 1 … 3328}. To represent one coefficient, we need 12 bits, so a straightforward representation takes 256 $\times$ 12 $=$ 3072 bits in length for each operand. Similarly, each polynomial in NewHope and qTesla algorithms contain 1024 coefficients and each coefficient is 16 bits in length, so 16384 (1024 $\times$ 16) bits for each operand is required. Similarly, the required operand length of multiplier functions in Crystals-Dilithium algorithm is 8192 bits. 

\textbf{Montgomery multiplier \cite{10.1007/3-540-48059-5_9}.} Montgomery multiplication is a faster way to get products and is often preferred in cryptographic applications \cite{article30}. To perform the multiplication, the input operands are first converted into Montgomery domain and then multiplication is performed on Montgomery representation. For complete mathematical formulations regarding the Montgomery algorithm, we redirect the reader to \cite{10.1007/3-540-48059-5_9}. The functions named \textit{mq\_poly\_montymul\_ntt()} and \textit{mq\_montymul()} are implemented in Falcon algorithm for polynomial multiplications using the Montgomery multiplication. The operand size in \textit{mq\_montymul()} function is 32 bits, while in \textit{mq\_poly\_montymul\_ntt()} function, 24576 bits are utilizes.

The required computational cost (in terms of clock cycles) of SBM, 2-Way KM, 4-Way TCM, NTT, and Montgomery multiplication methods, when ignoring coordinate/domain conversions, are \textit{m-1}, \textit{(m/2)-1}, \textit{(m/4)-1}, \textit{2m+2log\textsubscript{2}$^n$} and \textit{2m-1}, where \textit{m} is the operand length. Moreover, the lattice-based NIST PQC candidates perform reduction operations over prime and binary numbers/fields/rings. The number of clock cycles required for reduction routines can be a function of the multiplier architecture and the number of clock cycles required for each multiplication operation. We expand on this argument in Section \ref{sec:impl_mul}

\subsection{Hashing algorithms} \label{sec:Hash_Algos}
Apart from the memories and arithmetic operators, NIST PQC algorithms also make use of known cryptographic primitives. Different hash functions and block ciphers are employed for different purposes (e.g., AES and SHA are utilized for generating random numbers, secure hashing in key generations, encapsulation/decapsulation, encryption/decryption processes, as well as in message digests). In short, a hashing function transforms data (message string) of arbitrary size to fixed-size values. The values returned by a hash function as an output is termed as hash values, codes, digests, or hashes. In practice, a hash function may be considered to perform three functions, i.e., (1) convert variable length keys into fixed length, (2) scramble the bits of a key so that the resulting values are uniformly distributed over the key space, and (3) map key values into ones less than or equal to the size of the message. However, in NIST PQC algorithms, hash functions are frequently utilized for scrambling the bits of the key so that the resulting values are uniformly distributed over key space.

The variants of the SHA family of secure hash functions are SHA0, SHA1, SHA2, and SHA3. As shown in Table \ref{tab:table5}, the lattice-based PQC algorithms involved in the competition use a combination of both SHA2 and SHA3 functions. The first column in Table \ref{tab:table5} provides the name of algorithm along with reference model in parentheses, while the name of the hash function is presented in the second column. We have placed a dash ``--$"$ symbol in the entries of Table \ref{tab:table5} where neither hash nor AES algorithm is utilized. 

The SHA2 family consists of six hash functions, named as SHA-224, SHA-256, SHA-384, SHA-512, SHA-512/224 and SHA-512/256, while SHA3 family constitutes four hash functions, termed as SHA3-224, SHA3-256, SHA3-384, SHA3-512, and two extendable-output hash functions (XOF), named SHAKE-128 and SHAKE-256. The SHA2 and SHA3 functions and extendable-output hash functions that take two parameters (\textit{M, d}). The first parameter denotes the input message \textit{M} while the second parameter \textit{d} determines the output length in bits. The value of \textit{d} for all hash functions is fixed, while for XOF functions, SHAKE-128 and SHAKE-256, it depends on the user. The SHAKE-128 and SHAKE-256 functions can be used as customizable SHAKE (cSHAKE-128 and cSHAKE-256) with two additional inputs, i.e., function name bit string (\textit{N}) and a customization bit string (\textit{S}). Whenever, \textit{N} and \textit{S} are both empty strings then cSHAKE (\textit{M}, \textit{d}, \textit{N}, \textit{S}) is equivalent to SHAKE (\textit{M}, \textit{d}).

\begin{table}[h]
\begingroup
\setlength{\tabcolsep}{1.5pt} 
\renewcommand{\arraystretch}{1.5} 
\fontsize{8}{10}\selectfont

\begin{longtable}{|p{5cm}|p{8cm}|}
\caption{Hash algorithms utilized in the reference implementations of lattice-based PQC algorithms} \label{tab:table5} \\ 
\hline
\multirow{1}{*}{\parbox{5.0cm}{\textbf{Algorithm (reference model)}}} & \multirow{1}{*}{\parbox{7cm}{\textbf{Name of the hashing method utilized}}} \\ \hline

\multirow{1}{*}{NTRU-Prime (sntrup857)} & \multirow{1}{*}{SHA2-512} \\ \hline

\multirow{2}{*}{FrodoKEM (frodokem1344)} & \multirow{2}{*}{\parbox{8cm}{AES-128 (or) SHAKE-128 for matrix generation \\ SHAKE-128 (or) SHAKE-256 for key generation/encryption}} \\ 
& \\ \hline

\multirow{1}{*}{Saber (firesaber)} & SHAKE-128, SHA3-256 and SHA3-512 \\ \hline

\multirow{1}{*}{NTRU (hps4096821)} & \multirow{1}{*}{SHA3-256} \\ \hline

\multirow{1}{*}{ThreeBears (papabearephem)} & \multirow{1}{*}{cSHAKE-256} \\ \hline

\multirow{1}{*}{Round5 (r5nD-5pke-5d)} & \multirow{1}{*}{cSHAKE-256 and AES-256} \\ \hline

\multirow{1}{*}{Crystals-Dilithium (dilithium4-AES)} & \multirow{1}{*}{SHAKE-128 and SHAKE-256} \\ \hline

\multirow{1}{*}{Crystals-KYBER (kyber1024-90s)} & AES-256, SHA2-256, SHA2-512 and SHAKE-256 \\ \hline

\multirow{1}{*}{NewHope (newhope1024cca)} & \multirow{1}{*}{SHAKE-128 and SHAKE-256} \\ \hline

\multirow{1}{*}{LAC (lac256)} & - \\ \hline

\multirow{1}{*}{qTesla (qtesla-p-III)} & SHAKE-256, cSHAKE-128 and cSHAKE-256  \\ \hline

\multirow{1}{*}{Falcon (falcon1024)} & SHAKE-256 \\ \hline
\end{longtable}
\endgroup
\end{table}
In the text that follows, we describe the use of hash methods and block ciphers in selected NIST PQC algorithms. The required output lengths (in bits) for SHAKE and cSHAKE variants are also described below. 
\begin{itemize}
    \item \textbf{NTRU-Prime.} NTRU-Prime uses SHA2-512 for hashing a session key. In the reference implementation, only the first 256 bits of output generated by SHA-512 have been used, as the idea was to reduce the requirement for inclusion of extra bits at inputs (as used in reference \cite{CFD2013} to generate the public key) and makes it flexible to switch to another 256 bit hash function if desired. For mathematical descriptions, readers can redirect to sections 3.1 and 4.7 in reference \cite{CFD2013}.
    
    \item \textbf{FrodoKEM.} The FrodoKEM reference implementation comes in two flavours, either using AES or SHAKE. Naturally, the AES variants (FrodoKEM-640-AES, FrodoKEM-976-AES, and FrodoKEM-1344-AES) are more convenient for devices containing AES hardware acceleration such as AES-NI on Intel platforms. On the other hand, SHAKE variants (FrodoKEM-640-SHAKE, FrodoKEM-976-SHAKE, and FrodoKEM-1344-SHAKE) provide better performance in comparison with AES variants when no hardware acceleration is in place \cite{CFD2012}. To pseudorandomly generate a large public matrix, the AES variants of FrodoKEM use AES-128 while SHAKE variants uses SHAKE-128 (see algorithms 7--8 on page 16 of \cite{CFD2012}). For key generation and encryption processes, all variants of FrodoKEM use SHAKE-128 (or) SHAKE-256 (see algorithms 9--10 on page 17 of \cite{CFD2012}). Therefore, in our evaluations, we have used an instance of AES-128 for matrix generation and an instance of SHAKE-256 for key generation and encryption processes. Based on the parameters provided in \cite{CFD2012}, implementation of SHAKE takes an arbitrary length of string (0x5F||seed\textsubscript{SE}) as an input and generates an output length 344064 (in bits) for key generation (see line 4 of algorithm 9 in \cite{CFD2012}). In context of hardware implementations, the similar SHAKE core could be used for encryption process as it generates an output length smaller than the key generation. Hence, we have chosen an instance of SHAKE-256 to generate an output length 1024 (in bits) for both key generation and encryption processes. Using this instance to generate an hashing output over 344064 bits, 336 clock cycles are required (ratio of 344064 to 1024). 

    \item \textbf{Saber.} SHAKE-128, SHA3-256, and SHA3-512 are utilized in the reference implementation of Saber. SHAKE-128 is used for generation of pseudorandom matrix \textit{A} from a \textit{seed\textsubscript{A}} and for public and secret key pair generations (see algorithms 15--17 on page 22 \& 23 of \cite{CFD2022}). Different instances of SHAKE-128, in algorithms 15, 16 and 17 of \cite{CFD2022}, result in outputs of lengths 768, 6656, and 256 bits. For our evaluations, we have used an instance of SHAKE-128 with an output length 768 bits, as in hardware the similar resources of SHAKE-128 can be used to generate an output of lengths 6656 and 256 bits respectively. Similarly, the instances of SHA3-256 and SHA3-512 with an output byte string of lengths 32 and 64 are used in encapsulation and decapsulation processes of Saber (see algorithms 20--22 on pages 25 \& 26 of \cite{CFD2022}). 

    \item \textbf{NTRU.} In NTRU, SHA3-256 function with an output length of 256 bits is required for key encapsulation and decapsulation processes (see section 1.12 on page 19 of \cite{CFD2023}).

    \item \textbf{ThreeBears.} It uses cSHAKE-256 for multiple purposes such as to generate uniform and noise samplers (see algorithm 1 on page 18 of \cite{CFD2017}), keypair generation (see algorithm 5 on page 23 of \cite{CFD2017}), encapsulation (see algorithm 6 on page 25 of \cite{CFD2017}) and decapsulation (see algorithm 7 on page 26 of \cite{CFD2017}). Two different modifications have been considered in SHAKE-256 to develop cSHAKE-256. However, in the first one, they used an additional 1 byte $``$purpose$"$ \textit{p} to differentiate multiple purposes to each other and finally, they have introduced a zero byte between the purpose parameter and the purpose byte for word alignment in the incoming string. Based on the aforesaid modifications, the hash function the authors have used is cSHAKE256(pblock||[0.p]||data, 8.L, $``$ $"$, $``$ThreeBears$"$). For complete descriptions, readers can refer to section 2.4.1 on page 17 of \cite{CFD2017}. Therefore, the aforementioned cSHAKE function requires an output length 2496 bits for uniform and noise samplers while for remaining purposes an output length of 256 bits is required. Consequently, we have used an instance of aforementioned cSHAKE function with an output length 256 bits to evaluate the hardware resources.

    \item \textbf{Round5.} It uses a cSHAKE-256 and an instance of AES-256 for hashing purposes. cSHAKE-256 is used in the construction of the permutations while AES-256 is utilized as an alternative to generate random data. The cSHAKE-256 implementation takes a byte string as an input and outputs a hashed byte (see section 2.11.4 of \cite{CFD2021}). Moreover, cSHAKE-128 is used when the security parameter \textit{kappa} is equal to 128 (this is for lower security levels, i.e., \textit{SL\textsubscript{1}} and \textit{SL\textsubscript{3}}). Consequently, we have used only the cSHAKE-256 (with desired output length 256 bits) and an instance of AES-256 for our evaluations as we are considering only the highest security level \textit{SL\textsubscript{5}}.

    \item \textbf{Crystals-Dilithium.} Two XOF functions, i.e., SHAKE-128 and SHAKE-256, are required for matrix generation, signature signing, and verification procedures in Crystals-Dilithium. An instance of SHAKE-128 with an output length of 256 bits is used to generate a public matrix \textit{A} from a seed value to set a public key and the generated matrix is needed for both signature signing and its veriﬁcation (see lines 1 and 2 in Figure 4 on page 12 of \cite{CFD2025}). Similarly, an instance of SHAKE-256 with output length of 384 bits is used (see lines 7, 10, 12, and 27 on page 12 of the same figure). However, for complete descriptions, readers can turn to section 4.3 in reference \cite{CFD2025}.

    \item \textbf{Crystals-KYBER.} In the selected variant of Crystals-KYBER (kyber1024-90s), there are four hash functions utilized: an XOF using AES-256 in CTR mode, an \textit{H} function using SHA2-256, an  \textit{G} function using SHA2-512, a \textit{PRF(s, b)} function using AES-256 where \textit{s} is used as the key and \textit{b} is zero-padded to a 12-byte nonce, and finally a \textit{KDF} function that utilizes SHAKE-256 (see section 1.4 on page 11 of \cite{CFD2018}). These five functions are used in key generation, encryption, encapsulation and decapsulation processes (see algorithms 4, 5, and 7 to 9 on page 9 to 11 of \cite{CFD2018}). The requested output length in aforementioned functions is 256; therefore, we have used an instance of SHAKE-256 for our evaluations with an output length 256 bits.

    \item \textbf{NewHope.} SHAKE-128 and SHAKE-256 functions have been utilized with different inputs/outputs length in NewHope. For example, SHAKE-128 is used to generate the public parameters from a public seed \textit{state $\leftarrow$ SHAKE128Absorb(d)} function. It takes as input a byte array \textit{d} and results in a byte array of length 200 (\textit{state}) (see line 8 in algorithm 5 on page 9 of \cite{CFD2015}). In another case, SHAKE-128 is utilized as \textit{buf, state $\leftarrow$ SHAKE128Squeeze(j, state)} function where \textit{j} determines the amount of output blocks of SHAKE-128 to be produced and a state with 200 bytes length (see line 10 in algorithm 5 on page 9 of \cite{CFD2015}). It produces a byte array \textit{buf} of length 16 $\times$ \textit{j} and a byte array state of length 200 as an output. A SHAKE-256 function for different output lengths (512, 768 and 1024 bits) has been used to hash and extend the output of the random number generator in key generation, encapsulation and decapsulation functions (see algorithms 1, 4, 17--21 in reference \cite{CFD2015}). Therefore, we have used an instance each of SHAKE-128 and SHAKE-256 for output lengths 1600 (200 bytes$\times$8) and 1024 bits, respectively.

    \item \textbf{qTesla.} In the selected qtesla-p-III variant, an instance of SHAKE-256 is used for seed generation (see algorithm 9 in reference \cite{CFD2024}) and for hash functions \textit{G, H} (see algorithm 13 in reference \cite{CFD2024}). A cSHAKE-256 function is used to sample polynomial \textit{y} using \textit{GaussSampler} and \textit{ySampler} functions, respectively (see algorithms, 11 \& 12 in reference \cite{CFD2024}). cSHAKE-128 is used for generation of public polynomials using function \textit{GenA} (see algorithm 10 in reference \cite{CFD2024}) and for encoding purposes using function \textit{Enc} (see algorithm 14 in reference \cite{CFD2024}). The aforesaid functions have been utilized for different output lengths. Consequently, in our evaluation we have considered an instance of each XOF function: SHAKE-256, cSHAKE-128, and cSHAKE-256, all with identical output lengths of 256 bits.

    \item \textbf{Falcon.} SHAKE-256 is utilized in Falcon, where it takes an arbitrary length string as an input and produces 16 bits of hashed chunks as an output (see algorithm 3 at top of page 32 in reference \cite{CFD2026}). From an implementation point of view, it has been highlighted that it might be difficult to efficiently implement SHAKE-256 in a constant-time way (see section 4.5.2 for details on page 22 of reference \cite{CFD2026}). However, for efficient utilization of SHAKE-256, instead of 16, 64 bits can be extracted as an output \cite{CFD2026}. This was adopted in our evaluation as well.
\end{itemize}

Now that we have thoroughly analysed the reference implementations of selected NIST PQC candidates and identified all basic building blocks of interest, we can proceed with comparisons. The next section takes a look at how the studied algorithms would fare as ASIC accelerators, where we report area and power values for a given frequency of operation. Our approach has, without a doubt, many caveats. These are carefully discussed later in Section \ref{sec:5}.

\section{NIST PQC Candidates as ASIC Accelerators } \label{sec:4}
In this section, we provide actual implementation characteristics of the identified building blocks such that we can evaluate the lattice-based NIST PQC algorithms as ASIC accelerators. We make use of commercial memory compilers and a standard cell library, both designed for the same 65nm “low-power” technology. For the sake of comparison, while not revealing foundry privileged information, we clarify that the nominal supply voltage is 1.2V for both memories and logic. Without loss of generality, our analysis is performed only for the typical-typical corner (TT), i.e., P=1, V=1.2V, and T=25C.

In Section \ref{sec:4_memcalc}, we present our ROM and RAM data. In Section \ref{sec:impl_mul}, we provide the implementation results for selected SBM multiplier. Hashing functions and their implementation are given in Section \ref{sec:4.3}.

\subsection{Memory characteristics}  \label{sec:4_memcalc}
Area, max frequency, and power characteristics of each memory instance required by the many different algorithms are obtained from commercial memory compilers optimized for our 65nm technology of choice. The compilers output several files, including datasheets, LEF abstracts, LIB timing information, and GDSII layouts. The information presented in this section was extracted from over 50+ datasheets. 

Knowing the number of addresses and the size of each address is, unfortunately, not enough to estimate the characteristics of each memory instace. The memory compiler works under certain assumptions, so the user has to pick values that are within acceptable ranges. The most significant limitation comes from the choice of a column-mux ratio, which not only determines the aspect ratio of the actual memory, as well as bounds for address and data ranges. Thus, the valid range for \textit{p} (number of addresses) and \textit{q} (bits stored at each address) depends on the column-mux ratio. Considering the memory compilation for RAM, there are three available options to select column-mux ratio: 4, 8, and 16. Similarly, for ROMs, there are four available choices (4, 8, 16, and 32). While using different values of column-mux ratio does not affect the overall size of the generated memory, it does affect the performance. Without loss of generality, we have selected 8 as column-mux ratio for calculations of both ROM and RAM memory instances reported in this work.

\begin{table}[t]
\begingroup
\setlength{\tabcolsep}{1.5pt} 
\renewcommand{\arraystretch}{1.5} 
\fontsize{8}{10}\selectfont

\begin{longtable}{|p{3cm}|p{0.9cm}|p{0.9cm}|p{0.9cm}|p{0.9cm}|p{2.0cm}|p{1.5cm}|p{1.5cm}|}
\caption{Characteristics of ROMs generated using commercial memory compiler} \label{tab:table6} \\ 
\hline


\multirow{2}{*}{\parbox{3cm}{\textbf{Algorithm}}} &
\multicolumn{2}{l|}{\parbox{1.8cm}{\textbf{Input to compiler}}} &
\multicolumn{2}{l|}{\parbox{2.0cm}{\textbf{Memory\\dimensions\\(in $\mu$m)}}} &
\multirow{2}{*}{\parbox{2.0cm}{\textbf{Total area\\(in mm$^2$)\\=\textit{W} $\times$ \textit{H} $\times$ \textit{n}}}} & 
\multicolumn{2}{l|}{\parbox{3.0cm}{\textbf{Power consumption}}}\\\cline{2-5} \cline{7-8} 

{} & 
\textit{\textbf{p}} & 
\textit{\textbf{q}} & 
\textbf{\textit{W}} & 
\textbf{\textit{H}} &
{} &
\parbox{1.5cm}{\textbf{Static (\textit{$\mu$W})}} & 
\parbox{1.5cm}{\textbf{Dynamic\\(\textit{$\mu$W$\times$F})}}  \\ \hline








\multirow{1}{*}{\parbox{3cm}{Crystals-KYBER}} & {128} & {16} & {85.7} & {36.2} & {0.0062} & {0.175} & {5.541}\\ \hline

\multirow{1}{*}{\parbox{3cm}{NewHope}} & {1024} & {16} & {85.7} & {80.4} & {0.0276} & {0.310} & {8.455}\\ \hline

\multirow{2}{*}{\parbox{3cm}{LAC}} & {512} & {16} & {85.7} & {55.1} & {0.0094} & {0.232} & {7.0224}\\ \cline{2-8}
& {5120} & {32} & {128.7} & {308.3} & {0.0397} & {1.281} & {17.750} \\ \hline

\multirow{3}{*}{\parbox{3cm}{qTesla}} & {444} & {32} & {128.7} & {52.0} & {0.0067} & {0.266} & {11.553}\\ \cline{2-8}
&{224} & {64} & {214.6} & {40.9} & {0.0088} & {0.316} & {17.9650}\\ \cline{2-8}
&{2048} & {36} & {139.4} & {137.2} & {0.0383} & {0.600} & {17.089}\\ \hline

\multirow{11}{*}{\parbox{3cm}{Falcon}} & {540} & {64} & {214.6} & {56.7} & {0.0122} & {0.364} & {22.329}\\ \cline{2-8}
&{1080} & {16} & {85.7} & {89.8} & {0.0077} & {0.379} & {8.520}\\ \cline{2-8}
&{31} & {64} & {214.6} & {31.4} & {0.0135} & {0.288} & {15.190}\\ \cline{2-8}
&{27} & {64} & {214.6} & {31.4} & {0.0135} & {0.288} & {15.190}\\ \cline{2-8}
&{30} & {64} & {214.6} & {31.4} & {0.0135} & {0.288} & {15.190}\\ \cline{2-8}
&{1024} & {16} & {85.7} & {80.4} & {0.0138} & {0.309} & {8.455}\\ \cline{2-8}
&{32} & {16} & {85.7} & {31.4} & {0.0054} & {0.160} & {5.121}\\ \cline{2-8}
&{64} & {16} & {85.7} & {33.0} & {0.0057} & {0.165} & {5.272}\\ \cline{2-8}
&{1024} & {8} & {64.2} & {80.4} & {0.0103} & {0.309} & {8.455}\\ \cline{2-8}
&{256} & {8} & {64.2} & {42.5} & {0.0055} & {0.174} & {3.988}\\ \cline{2-8}
&{512} & {8} & {64.2} & {55.1} & {0.0071} & {0.232} & {7.022}\\ \hline
\end{longtable}
\endgroup
\end{table}
The ROM and RAM memory compilers used in this work are very similar. The main difference is that we have to define the data content for the ROM compiler because it does not initialize data. Therefore, we had to create data initialization files for each generated ROM memory instances. To this end, we have prepared a MATLAB script to make sure that the ROMs are initialized with a proper ratio of zeros and ones, otherwise the power estimation would be skewed. The output generated by MATLAB has been converted into an 'Intel format' using a Perl script. The converted file is then passed as an input to the memory compiler for the actual compilation.

The values collected from the generated compiled memories are shown in Tables \ref{tab:table6} and \ref{tab:table7}. The first column in Tables \ref{tab:table6} and \ref{tab:table7} lists the related algorithm while the second column presents numerical values for \textit{p} (memory addresses) and \textit{q} (number of bits at each address) that were provided as inputs to memory compiler. Due to certain limitations in the tool, as discussed earlier, the present form of values of \textit{p} and \textit{q} (shown in second column of Tables \ref{tab:table6} and \ref{tab:table7}) could not be used directly as an input to the commercial memory compiler. Therefore, we have rounded off the values of \textit{p} and \textit{q} to make it possible to use as an input to the memory compiler. The sizes of the ROM and RAM instances, given by \textit{W} (width of the memory cell) and \textit{H} (height of the memory cell), are shown in column three of Tables \ref{tab:table6} and \ref{tab:table7}. For \textit{n} number of required memory instances, the total area is calculated by product of width of the memory cell, height of the memory cell and the number of required memory instances (\textit{n}), as shown in column four of Tables \ref{tab:table6} and \ref{tab:table7}. Finally, power consumption values, both static (in \textit{$\mu$W}) and dynamic (\textit{$\mu$W$\times$F}), are provided in column five of Tables \ref{tab:table6} and \ref{tab:table7}. All generated memories are single port and perform one operation per clock cycle.

\begingroup
\setlength{\tabcolsep}{1.5pt} 
\renewcommand{\arraystretch}{1.5} 
\fontsize{8}{10}\selectfont
\begin{longtable}{|p{3cm}|p{0.9cm}|p{0.9cm}|p{0.9cm}|p{0.9cm}|p{2.0cm}|p{1cm}|p{1.5cm}|p{1.5cm}|}
\caption{Characteristics of RAMs generated using commercial memory compiler} \label{tab:table7} \\ 
\hline


\multirow{2}{*}{\parbox{3cm}{\textbf{Algorithm}}} &
\multicolumn{2}{l|}{\parbox{1.8cm}{\textbf{Inputs to compiler}}} &
\multicolumn{2}{l|}{\parbox{2.0cm}{\textbf{Memory\\dimensions\\(in $\mu$m)}}} &
\multirow{2}{*}{\parbox{2.0cm}{\textbf{Total area\\(in mm$^2$)\\=\textit{W} $\times$ \textit{H} $\times$ \textit{n}}}} & 
\multicolumn{2}{l|}{\parbox{3.0cm}{\textbf{Power consumption}}}\\\cline{2-5} \cline{7-8} 

{} & 
\textit{\textbf{p}} & 
\textit{\textbf{q}} & 
\textbf{\textit{W}} & 
\textbf{\textit{H}} &
{} &
\parbox{1.5cm}{\vspace{1mm}\textbf{Static (\textit{$\mu$W})}\vspace{1mm}} & 
\parbox{1.5cm}{\textbf{Dynamic\\(\textit{$\mu$W$\times$F})}}  \\ \hline

\multirow{2}{*}{\parbox{3.0cm}{NTRU-Prime}} & {256} & {8} & {112.8} & {160.5} & {0.1087} & {0.331} & {8.307}\\ \cline{2-9}
&{24} & {64} & {79.2} & {161.5} & {0.0768} & {0.331} & {5.488}\\ \hline 

\multirow{2}{*}{\parbox{3.0cm}{FrodoKEM}} & {10752} & {16} & {311.1} & {59.1} & {0.0184} & {0.2136} & {22.656}\\\cline{2-9}
& {64} & {16} & {109.5} & {59.1} & {0.0065} & {0.084} & {6.613} \\ \hline

\multirow{10}{*}{\parbox{3cm}{Saber}} & {32} & {8} & {75.9} & {55.1} & 0.0042 & {0.056} & {3.832} \\ \cline{2-9}
&{32} & {16} & {109.5} & {55.1} & 0.0121 & {0.112} & {6.452} \\ \cline{2-9}
&{128} & {8} & {75.9} & {67.1} & 0.0051 & {0.076} & {4.149} \\ \cline{2-9}
&{128} & {16} & {109.5} & {67.1} & 0.0174 & {0.103} & {6.932} \\ \cline{2-9}
&{64} & {8} & {75.9} & {59.1} & 0.0045 & {0.062} & {3.938} \\ \cline{2-9}
&{64} & {16} & {109.5} & {59.1} & 0.0065 & {0.084} & {6.613} \\ \cline{2-9}
&{4} & {512} & {647.1} & {55.1} & 0.0357 & {0.376} & {13.083} \\ \cline{2-9}
&{4} & {1024} & {647.1} & {55.1} & 0.0357 & {0.376} & {5.758} \\ \cline{2-9} 
&{4} & {256} & {647.1} & {55.1} & 0.0357 & {0.376} & {7.156} \\ \cline{2-9}
&{4} & {512} & {647.1} & {55.1} & 0.0357 & {0.376} & {13.083} \\ \hline

\multirow{1}{*}{\parbox{3cm}{NTRU}} & {821} & {16} & {112.8} & {156.5} & {0.2473} & {0.319} & {82.280} \\ \hline

\multirow{5}{*}{\parbox{2.6cm}{ThreeBears}} & {40} & {8} & {75.9} & {56.1} & {0.0043} & {0.057} & {3.859}\\ \cline{2-9}
{} & {1584} & {8} & {79.2} & {253.9} & {0.0201} & {0.399} & {6.772}\\ \cline{2-9}
{} & {1697} & {8} & {79.2} & {268.9} & {0.0213} & {0.426} & {6.978}\\ \cline{2-9}
{} & {24} & {8} & {75.9} & {55.1} & {0.0042} & {0.056} & {3.832}\\ \cline{2-9}
{} & {32} & {8} & {75.9} & {55.1} & {0.0042} & {0.056} & {3.832}\\ \hline

\multirow{2}{*}{\parbox{3cm}{Round5}} & {16} & {8} & {75.9} & {55.1} & {0.0042} & {0.056} & {3.832}\\ \cline{2-9}
&{32} & {8} & {75.9} & {55.1} & {0.0042} & {0.056} & {3.832} \\ \hline

\multirow{1}{*}{\parbox{3cm}{Crystals-Dilithium}} & {256} & {32} & {176.7} & {83.1} & {0.0441} & {0.216} & {12.762}\\ \hline

\multirow{2}{*}{\parbox{3cm}{Crystals-KYBER}} & {256} & {16} & {109.5} & {83.1} & {0.0729} & {0.141} & {7.156}\\ \cline{2-9}
{} & {128} & {16} & {109.5} & {67.1} & {0.0074} & {0.103} & {6.932}\\ \hline

\multirow{1}{*}{\parbox{3cm}{NewHope}} & {1024} & {16} & {112.8} & {181.5} & {0.1639} & {0.379} & {8.624}\\ \hline

\multirow{3}{*}{\parbox{3cm}{LAC}} & {2080} & {8} & {183.7} & {32.8} & {0.0121} & {70.628} & {7.563}\\ \cline{2-9}
&{1056} & {8} & {79.2} & {186.3} & {0.0148} & {0.285} & {5.874} \\ \cline{2-9} \pagebreak \cline{2-9}
&{1024} & {8} & {79.2} & {181.5} & {0.0144} & {0.279} & {5.758} \\ \hline

\multirow{5}{*}{\parbox{3cm}{qTesla}} & {2048} & {8} & {79.2} & {312.7} & {0.0248} & {0.319} & {11.553}\\ \cline{2-9}
&{9600} & {32} & {180.0} & {312.7} & {0.2815} & {1.050} & {69.676}\\ \cline{2-9}
&{10240} & {32} & {180.0} & {312.7} & {0.2815} & {1.050} & {87.096}\\ \cline{2-9}
&{1408} & {32} & {180.0} & {231.1} & {0.0416} & {0.756} & {15.661} \\ \cline{2-9} 
&{2048} & {64} & {314.4} & {312.7} & {0.3934} & {1.784} & {30.560}\\ \hline

\multirow{2}{*}{\parbox{3cm}{Falcon}} & {1024} & {16} & {112.8} & {181.5} & {0.0205} & {0.379} & {8.624}\\ \cline{2-9}
&{521} & {32} & {180.0} & {118.7} & {0.1287} & {0.454} & {13.83}\\ \hline
\end{longtable}
\endgroup
\subsection{Implementation of identified multipliers} \label{sec:impl_mul}
It is important to emphasize that there are many options available to implement the multiplication operations required by the studied algorithms. For this reason, we provide a thorough discussion about multiplier architectures in Appendix \ref{appendix:A}. There, we compare SBM, 2-Way KM, 3-Way TCM, and 4-Way TCM over a range of large operand sizes. The outcome of this discussion is that the SBM is advantageous as it leads to low area, reasonable frequency of operation, and, most importantly, the lowest power consumption of the compared architectures. This characteristic, even if it comes at a cost in latency, led us to select the SBM as the multiplier of choice for comparing all lattice-based candidates fairly. In order to provide a fair comparison in terms of clock frequency, we make use of 500MHz as our frequency of choice in the experiments that follow\footnote{Synthesis is performed with high area and power effort on 65nm technology}.

We have presented the implementation results using SBM multiplier for each identified operand size in Table \ref{tab:table8}. The implemented SBM multiplier takes two inputs \textit{OP\textsubscript{1}} and \textit{OP\textsubscript{2}} with corresponding size \textit{m} and \textit{n} bits, respectively. It results in a product with size $\textit{s} = \textit{m}+\textit{n}-1$ bits. Therefore, it is essential to perform reduction to achieve a product length that matches the input length. 

One approach to perform reduction is to perform bitwise XOR operation over first \textit{m} bits of \textit{s} with cyclic shifts to left on the remaining \textit{n} bits of \textit{s}, one by one. This process can be repeated until all \textit{n} bits of \textit{s} are processed. This approach is a good fit for use in conjunction with the SBM multiplier since it already implements a shifter of size \textit{s}. Thus, a unified architecture for reduction and multiplication can be sought. While we have previously said that the SBM multiplier takes \textit{m-1} cycles, a unified multiplication/reduction unit takes \textit{2(m-1)} cycles to compute.

Our results for the implemented multipliers are shown in Table \ref{tab:table8}, where the first column shows the name of the particular algorithm and the length of input operands in parentheses (i.e., \textit{OP\textsubscript{1} $\times$ OP\textsubscript{2}}). The number of combinational and the sequential cells are provided in the second and third column of Table \ref{tab:table8}, respectively. The reported area (in \textit{$mm^2$}) is provided in the fourth column. Finally, columns five and six provide the dynamic power (in \textit{$\mu$W}) and leakage power (also in \textit{$\mu$W}) for multiplication operation. 

\begin{table}[t]
\begingroup
\setlength{\tabcolsep}{1.5pt} 
\renewcommand{\arraystretch}{1.5} 
\fontsize{8}{10}\selectfont

\begin{longtable}{|m{4.2cm}|m{2.3cm}|m{1.8cm}|m{1.3cm}|m{1.5cm}|m{1.5cm}|}
\caption{Implementation results for SBM multiplier using a 65nm standard cell library. Target frequency is 500MHz.} \label{tab:table8} \\ 
\hline 

\multirow{3}{*}{\parbox{4.5cm}{\textbf{Algorithms\\(operand sizes in bits)}}} & 
\multirow{3}{*}{\parbox{2.2cm}{\textbf{Combinational cells}}} & \multirow{3}{*}{\parbox{1.6cm}{\textbf{Sequential\\cells}}} & \multirow{3}{*}{\parbox{1.5cm}{\textbf{Area (\textit{mm$^2$})}}} & \multirow{3}{*}{\parbox{1.5cm}{\textbf{Dynamic power (\textit{$\mu$W})}}} & 
\multirow{3}{*}{\parbox{1.5cm}{\textbf{Leakage power (\textit{$\mu$W})}}} \\
{} & {} & {} & {} & {} & {} \\
{} & {} & {} & {} & {} & {} \\ \hline

\multirow{1}{*}{\parbox{4.5cm}{NTRU-Prime (6088$\times$6088)}} & \multirow{1}{*}{\parbox{2cm}{435073}} & 
\multirow{1}{*}{\parbox{2cm}{12189}} & 
\multirow{1}{*}{\parbox{1.5cm}{0.8389}} & 
\multirow{1}{*}{\parbox{1.5cm}{100510.6}} & 
\multirow{1}{*}{\parbox{1.5cm}{125.3402}} \\ \hline

\multirow{1}{*}{\parbox{4.5cm}{NTRU-Prime (12176$\times$6088)}} & \multirow{1}{*}{\parbox{2cm}{750152}} & 
\multirow{1}{*}{\parbox{2cm}{18278}} & 
\multirow{1}{*}{\parbox{1.5cm}{1.4124}} & 
\multirow{1}{*}{\parbox{1.5cm}{144073.8}} & 
\multirow{1}{*}{\parbox{1.5cm}{202.0082}} \\ \hline

\multirow{1}{*}{\parbox{4.5cm}{FrodoKEM\footnote{The area and power values reported for FrodoKEM are estimated instead of synthesized. Unfortunately a multiplier of this size is too challenging for synthesis to handle.} (172032$\times$172032)}} & 
{-} & 
{2329421} & 
{22.1505} &  
{51177810.} & 
{629.637}  \\ \hline

\multirow{1}{*}{\parbox{4.5cm}{Saber (4096$\times$4096)}} & \multirow{1}{*}{\parbox{2cm}{206941}}  & 
\multirow{1}{*}{8205} & 
\multirow{1}{*}{0.4599} & 
\multirow{1}{*}{66433.2} & 
\multirow{1}{*}{42.4874} \\ \hline

\multirow{1}{*}{\parbox{4.5cm}{NTRU (11216$\times$11216)}} & 
{821229} & 
{22446} & 
{1.5602} & 
{163509.7} & 
{207.3250} \\ \hline

\multirow{1}{*}{\parbox{4.5cm}{ThreeBears (3120$\times$3120)}} & 
{141422}  & 
\multirow{1}{*}{6252} & 
\multirow{1}{*}{0.3192} & 
\multirow{1}{*}{58829.7} & 
\multirow{1}{*}{30.8762} \\ \hline

\multirow{1}{*}{\parbox{4.5cm}{Round5 (15136$\times$6208)}} & 
{902623}  & 
{21358} &
{1.6785} &
{164367.2} & 
{230.1910} \\ \hline

\multirow{1}{*}{\parbox{4.5cm}{Crystals-Dilithium (8192$\times$8192)}} & \multirow{1}{*}{\parbox{2cm}{592155}} & 
\multirow{1}{*}{\parbox{2cm}{16398}} & 
\multirow{1}{*}{\parbox{1.5cm}{1.1251}} & 
\multirow{1}{*}{\parbox{1.5cm}{123343.3}} & 
\multirow{1}{*}{\parbox{1.5cm}{139.4685}} \\ \hline

\multirow{1}{*}{\parbox{4.5cm}{Crystals-KYBER (3072$\times$3072)}} & \multirow{1}{*}{\parbox{2cm}{131973}} & 
\multirow{1}{*}{\parbox{2cm}{6156}} & 
\multirow{1}{*}{\parbox{1.5cm}{0.3069}} & 
\multirow{1}{*}{\parbox{1.5cm}{51800.5}} & 
\multirow{1}{*}{\parbox{1.5cm}{31.9299}} \\ \hline

\multirow{1}{*}{\parbox{4.5cm}{NewHope (16384$\times$16384)}} & \multirow{1}{*}{\parbox{2cm}{1302689}}  & 
\multirow{1}{*}{32783} & 
\multirow{1}{*}{2.4760} & 
\multirow{1}{*}{230704.9} & 
\multirow{1}{*}{446.6865} \\ \hline

\multirow{1}{*}{\parbox{2.5cm}{LAC (8192$\times$8192)}} & 
{592155} & 
\multirow{1}{*}{\parbox{2cm}{16398}} & 
\multirow{1}{*}{\parbox{1.5cm}{1.1251}} & 
\multirow{1}{*}{\parbox{1.5cm}{123343.3}} & 
\multirow{1}{*}{\parbox{1.5cm}{139.4685}} \\ \hline

\multirow{1}{*}{\parbox{4.5cm}{qTesla (16384$\times$16384)}} & \multirow{1}{*}{\parbox{2cm}{1302689}}  & 
\multirow{1}{*}{32783} & 
\multirow{1}{*}{2.4760} & 
\multirow{1}{*}{230704.9} & 
\multirow{1}{*}{446.6865} \\ \hline

\multirow{1}{*}{\parbox{4.5cm}{Falcon (32$\times$32)}} & \multirow{1}{*}{\parbox{2cm}{1001}}  & 
\multirow{1}{*}{70} & 
\multirow{1}{*}{0.0024} & 
\multirow{1}{*}{581.5} & 
\multirow{1}{*}{0.2499} \\ \hline

\multirow{1}{*}{\parbox{4.5cm}{Falcon (24576$\times$24576)}} & \multirow{1}{*}{\parbox{2cm}{2926129}}  & 
\multirow{1}{*}{49167} & 
\multirow{1}{*}{5.4670} & 
\multirow{1}{*}{327836.5} & 
\multirow{1}{*}{1410.9000} \\ \hline
\end{longtable}
\endgroup
\end{table}
As shown in column three of Table \ref{tab:table8}, the number of flip flops for SBM multiplier are roughly 2 times the length of input operands. A handful of additional flops are required for controlling the shift \& add operation and for setting its termination condition. The overall area (in $mm^2$) is reported in column four of Table \ref{tab:table8}, while power values are given in in columns five and six.

As summarized, we have achieved the multiplication results using SBM multiplier architecture by generating partial products for operand lengths as large as 24576$\times$24576 bits (for Falcon algorithm). There are numerous research practices where input sizes are much more modest. For instance, \cite{harvey_2013}, \cite{Roy2019}, \cite{Rafferty2017} employ operand sizes as large as 4096 bits. However, despite the SBM multiplier, we have shown implementation results for operand size of 4096 bits over 2-Way KM, 3-Way TCM, and 4-Way TCM multipliers, which is comparable with state-of-the-art implementations \cite{harvey_2013}, \cite{Roy2019}, \cite{Rafferty2017}. These results are shown in Appendix \ref{appendix:A}.
\subsection{Implementation of identified hash algorithms}  \label{sec:4.3}
The lattice-based NIST PQC candidates use different hash functions with different input and output lengths. Previously, in Section \ref{sec:Hash_Algos} of this work, we have described the required lengths for input and output parameters of identified hash functions. This section deals with the implementation of identified hash functions. Therefore, we have developed our own RTL cores in Verilog (HDL) for the identified hash and XOF functions. In one clock cycle, each developed core takes a 64 bits of a message as an input and results the desired hash value as an output (as described in Section \ref{sec:Hash_Algos}). Furthermore, in the developed cores, \textit{M\textsubscript{length}/64} clock cycles are required to read the input message string when \textit{M\textsubscript{length}} is the message length. The total number of clock cycles to generate one hash value is depends on the number of state rounds in the hash and XOF functions. Different number of rounds are for different hash and XOF functions. However, sum of clock cycles for \textit{M\textsubscript{length}/64} and for number of rounds determines the total clock cycles to generate a hash value over an arbitrary length of message. For each hash and XOF function, the behavioral simulations of each developed core is verified by providing an empty `` $"$ message string as an input and the resultant hash value achieved in 256 and 512 bits in length is compared with the corresponding hash values (test vectors for verification's -- provided by the NIST -- available at \cite{hash_TV}). The implementation results over 500MHz on commercial 65nm standard library are provided in Table \ref{tab:table9}. The name of the NIST PQC candidate is provided in the first column of Table \ref{tab:table9}, while the corresponding utilized hash/cipher is shown in column two. Required area (in $\mu m^2$) is shown in column three, while columns four and five provide values for dynamic (in $\mu W$) and leakage power (in $\mu W$), respectively. Finally, we have used a dash symbol $``-"$ in Table \ref{tab:table9} where neither hash nor AES block cipher is required.


\begingroup
\setlength{\tabcolsep}{1.5pt} 
\renewcommand{\arraystretch}{1.5} 
\fontsize{8}{10}\selectfont

\begin{longtable}{|p{3.5cm}|p{3.2cm}|p{2.0cm}|p{2.0cm}|p{2.0cm}|}
\caption{Implementation of identified hash algorithms using 65nm standard cell library. Target frequency is 500MHz.} \label{tab:table9} \\ 
\hline 

\multirow{2}{*}{\parbox{3.5cm}{\textbf{Algorithms}}} & 
\multirow{2}{*}{\parbox{3.2cm}{\textbf{Hash functions}}} & \multirow{2}{*}{\parbox{2.0cm}{\textbf{Area (\textit{mm$^2$})}}} & \multirow{2}{*}{\parbox{2.0cm}{\textbf{Dynamic power (\textit{$\mu$W})}}} & \multirow{2}{*}{\parbox{2.0cm}{\textbf{Leakage power (\textit{$\mu$W})}}} \\
{} & {} & {} & {} & {} \\ \hline

\multirow{1}{*}{\parbox{2.8cm}{NTRU-Prime}} & 
\multirow{1}{*}{\parbox{1.8cm}{SHA2-512}} & 
\multirow{1}{*}{\parbox{1.7cm}{0.0732}} & 
\multirow{1}{*}{18003.4} & 
\multirow{1}{*}{1.6227}  \\ \hline

\multirow{2}{*}{\parbox{2.8cm}{FrodoKEM}} &
\parbox{1.8cm}{AES-128} & 
\parbox{1.7cm}{0.0225} & 
6415.5 & 
0.3062  \\ \cline{2-5} 
{} &
\parbox{1.8cm}{SHAKE-256} &
\parbox{1.7cm}{0.1056} &
18568.4 & 
3.6235  \\ \hline

\multirow{3}{*}{\parbox{2.8cm}{Saber}} & 
\parbox{1.8cm}{SHAKE-128} & 
0.1101 & 
19379.2 & 
3.1528  \\ \cline{2-5} 

{} &
\parbox{1.8cm}{SHA3-256} &
0.1062 & 
18568.1 & 
4.2955  \\ \cline{2-5} 

{} &
\parbox{1.8cm}{SHA3-512} &
\parbox{1.7cm}{0.0984} &
15927.1 & 
3.2830  \\ \hline

\multirow{1}{*}{\parbox{2.8cm}{NTRU}} & 
\parbox{1.8cm}{SHA3-256} & 
0.1062 & 
18568.1 & 
4.2955  \\  \hline

\multirow{1}{*}{\parbox{2.8cm}{ThreeBears}} & 
\multirow{1}{*}{\parbox{1.8cm}{cSHAKE-256}} & 
\multirow{1}{*}{\parbox{1.7cm}{0.1055}} & 
\multirow{1}{*}{18568.4} & 
\multirow{1}{*}{3.6235}  \\ \hline

\multirow{2}{*}{\parbox{2.8cm}{Round5}} & 
\parbox{1.8cm}{cSHAKE-256} & 
\multirow{1}{*}{\parbox{1.7cm}{0.1055}} & 
\multirow{1}{*}{18568.4} & 
\multirow{1}{*}{3.6235}  \\ \cline{2-5} 

{} &
\parbox{1.8cm}{AES-256} &
\parbox{1.7cm}{0.0395} &
13562.1 & 
0.4472  \\ \hline

\multirow{2}{*}{\parbox{2.8cm}{Crystals-Dilithium}} & 
\parbox{1.8cm}{SHAKE-128} & 
0.1103 & 
19649.9 & 
4.4626   \\ \cline{2-5} 
{} &
\parbox{1.8cm}{SHAKE-256} &
\parbox{1.7cm}{0.1056} &
18556.2 & 
3.5285 \\ \hline

\multirow{4}{*}{\parbox{2.8cm}{Crystals-KYBER}} & 
\parbox{1.8cm}{AES-256} & 
\parbox{1.7cm}{0.0395} &
13562.1 & 
0.4472   \\ \cline{2-5} 

{} &
\parbox{1.8cm}{SHA2-256} &
\parbox{1.7cm}{0.0362} &
8881.4 & 
0.4671  \\ \cline{2-5} 

{} &
\parbox{2.6cm}{SHA2-512} &
\multirow{1}{*}{\parbox{1.7cm}{0.0732}} & 
\multirow{1}{*}{18003.4} & 
\multirow{1}{*}{1.6227}   \\ \cline{2-5} 

{} &
\parbox{1.8cm}{SHAKE-256} &
\parbox{1.7cm}{0.1055} &
18568.4 & 
3.6235  \\ \hline

\multirow{2}{*}{\parbox{2.8cm}{NewHope}} & 
\parbox{1.8cm}{SHAKE-128} & 
0.1103 & 
19649.9 & 
4.4626    \\ \cline{2-5} 

{} &
\parbox{1.8cm}{SHAKE-256} &
\parbox{1.7cm}{0.1055} &
18555.9 & 
3.4193 \\ \hline

\multirow{1}{*}{\parbox{2.5cm}{LAC}} & 
\multirow{1}{*}{\parbox{2.6cm}{--}} & 
\multirow{1}{*}{\parbox{1.5cm}{--}} & 
\multirow{1}{*}{--} & 
\multirow{1}{*}{--}  \\ \hline

\multirow{3}{*}{\parbox{2.5cm}{qTesla}} & 
\parbox{2.6cm}{SHAKE-256} & 
\parbox{1.7cm}{0.1056} &
18568.4 & 
3.6235  \\ \cline{2-5} 

{} &
\parbox{2.6cm}{cSHAKE-128} &
\parbox{1.5cm}{0.1103} &
19649.9 & 
4.4626  \\ \cline{2-5} 

{} &
\parbox{2.6cm}{cSHAKE-256} &
\multirow{1}{*}{\parbox{1.7cm}{0.1055}} & 
\multirow{1}{*}{18568.4} & 
\multirow{1}{*}{3.6235}  \\ \hline

\multirow{1}{*}{\parbox{2.8cm}{Falcon}} & 
\multirow{1}{*}{\parbox{1.8cm}{SHAKE-256}} & 
\parbox{1.7cm}{0.1056} &
18559.8 & 
3.4941  \\ \hline
\end{longtable}
\endgroup

In the text that follows, we describe the  hash and XOF functions that we have implemented in RTL, as well as their parameters. Moreover, an open source RTL core for AES block cipher is also used in this work and discussed below.

\textbf{SHA2 cores.} The developed cores for SHA2-256 and SHA2-512 generate a hash value with a 256-bit or 512-bit output, respectively. The internal state length is set to 256 bits (8$\times$32) for SHA2-256 while for SHA2-512 it is set as 512 bits (8$\times$64). The block sizes are 512 and 1024 bits, respectively. Finally, the number of required rounds to generate one hash value are 64 for SHA2-256 and 80 for SHA2-512. The developed RTL core for SHA2-256 is only used in Crystals-KYBER algorithm. Similarly, the developed RTL core for SHA2-512 is required in NTRU-Prime and Crystals-KYBER algorithms. 

\textbf{SHA3 cores.} The cores for SHA3-256 and SHA3-512 are developed based on the function keccak[c](M||01, d), where c denotes the capacity (c=512 for SHA3-256 and c=1024 for SHA3-512), M represents the input message, and d represents the output length in bits (d=256 for SHA3-256 and d=512 for SHA3-512). In both cores, internal state length is set to 1600 bits (5$\times$5$\times$64). The used block size for SHA3-256 is 1088 bits and for SHA3-512 is 576 bits. Finally, to generate the digest message, the number of required rounds for SHA3-256 and SHA3-512 are 24. The developed core for SHA3-256 is used in NTRU and Saber algorithms while the SHA3-512 core is only used in Saber algorithm. 
 
\textbf{SHAKE cores.} We have developed two SHAKE cores, namely SHAKE-128 and SHAKE-256, which are based on the function keccak[c](M||1111, d), where the capacity c is 256 for SHAKE-128 and 512 for SHAKE-256. The internal state length is set to 1600 bits (5$\times$5$\times$64) for both cores. The blocks size for SHAKE-128 is 1344 bits and for SHAKE-256 is 1088 bits. To generate the digest message, the number of required rounds for SHAKE-128 and SHAKE-256 are 24. The SHAKE-128 core is required in Crystals-Dilithium, Saber, and NewHope algorithms, where the desired output lengths are 256, 768, and 1600 bits, respectively. Similarly, the SHAKE-256 core is used in Falcon, FrodoKEM, Crystals-Dilithium, Crystals-KYBER, NewHope, and qTesla algorithms. The output lengths are 64, 1024, 384, 256, 1024 and 256 bits, respectively.

\textbf{cSHAKE cores.} Our cSHAKE-128 and cSHAKE-256 cores are developed based on the function keccak[c](bytepad(encode\_string(N)||encode\_string(S), R)||M||00, d), where the capacity c is 256 for cSHAKE-128 and 512 for cSHAKE-256, N is the function name bit string, S is the customization bit string, R represents the rate (in bytes). Our cSHAKE cores use identical lengths for internal state length and block sizes as in the SHAKE cores. A digest message is also generated after 24 rounds. The cSHAKE-128 core is only required in qTesla algorithm (output lengths of 256 bits). The cSHAKE-256 core is used in ThreeBears, Round5, and qTesla algorithms, also with an output length of 256 bits.

\textbf{AES cores.} We have selected an open source RTL core (for 128 and 256 bits) written in Verilog (HDL) for AES algorithm from reference \cite{AES_CORE_128} and \cite{AES_CORE}, respectively. Consequently, the AES core for 128 bits is required in Frodo-KEM algorithm while the AES core for 256 bits is required in Crystals-KYBER and Round5 algorithms, respectively. The AES core of \cite{AES_CORE_128} for 128 requires 12 clock cycles to perform complete encryption. Similarly, the AES core mentioned in \cite{AES_CORE} need 46 clock cycles to achieve desired output.

\section{Evaluation of NIST PQC Algorithms as Hardware Accelerators} \label{sec:5}

In this section, we treat the algorithms as accelerators completely described as specialized hardware -- there are no processor/software components. Therefore, for each accelerator, we provide the aggregated results that take into account all the identified building blocks. Our results are given for area and power, while the target frequency is set at 500MHz for the sake of comparison. We emphasize that complete implementations of NIST lattice-based PQC algorithms are not described in this work for the reason that we have not accounted for the “glue logic” that gives meaning to each algorithm. 

Area and power figures for each accelerator are calculated by using Eq. \ref{eq:area} and Eq. \ref{eq:power}, respectively, in which we sum the contributions of each building block\footnote{NTRU-Prime and Falcon employ more than one multiplier. We sum the contributions of both multipliers for each algorithm. For NTRU-Prime in particular, this might not be the optimal design choice since both multipliers take at least one input of 6088 bits, meaning that resources can be easily shared.}.

\begin{equation} \label{eq:area}
Area = area\,of(\sum{ROM}+\sum{RAM}+MULT+\sum{HASH})
\end{equation}

\begin{equation} \label{eq:power}
Power = dynamic\,power\,of (\sum{ROM}+\sum{RAM}+MULT+\sum{HASH})
\end{equation}

\begin{figure}[ht]
\centering \footnotesize
\includegraphics[width=5in]{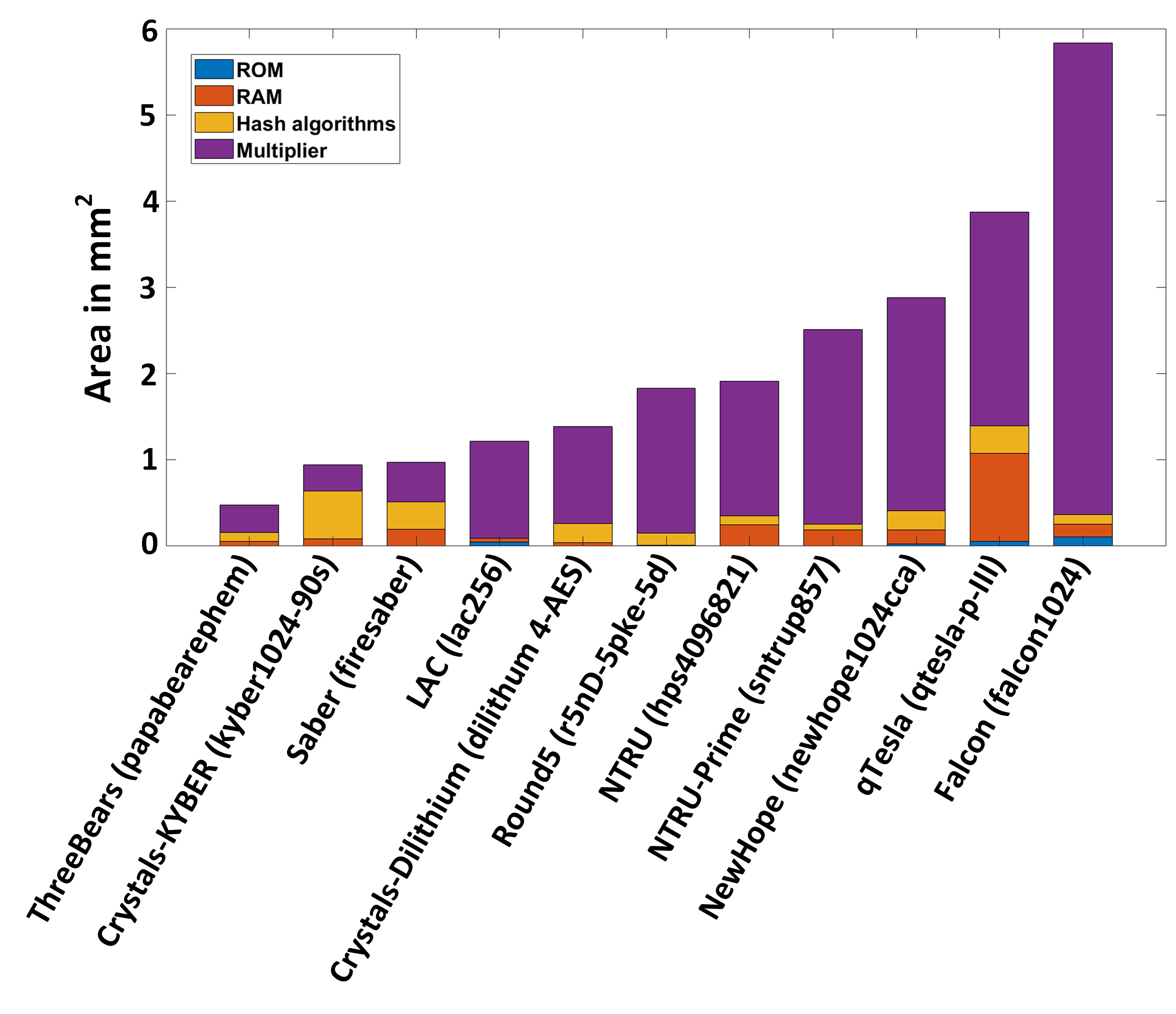}
\caption{Total area of the studied NIST lattice-based PQC algorithms, ordered.}
\centering
\label{fig:figure3}
\end{figure}

\begin{figure}[ht]
\centering \footnotesize
\includegraphics[width=5in]{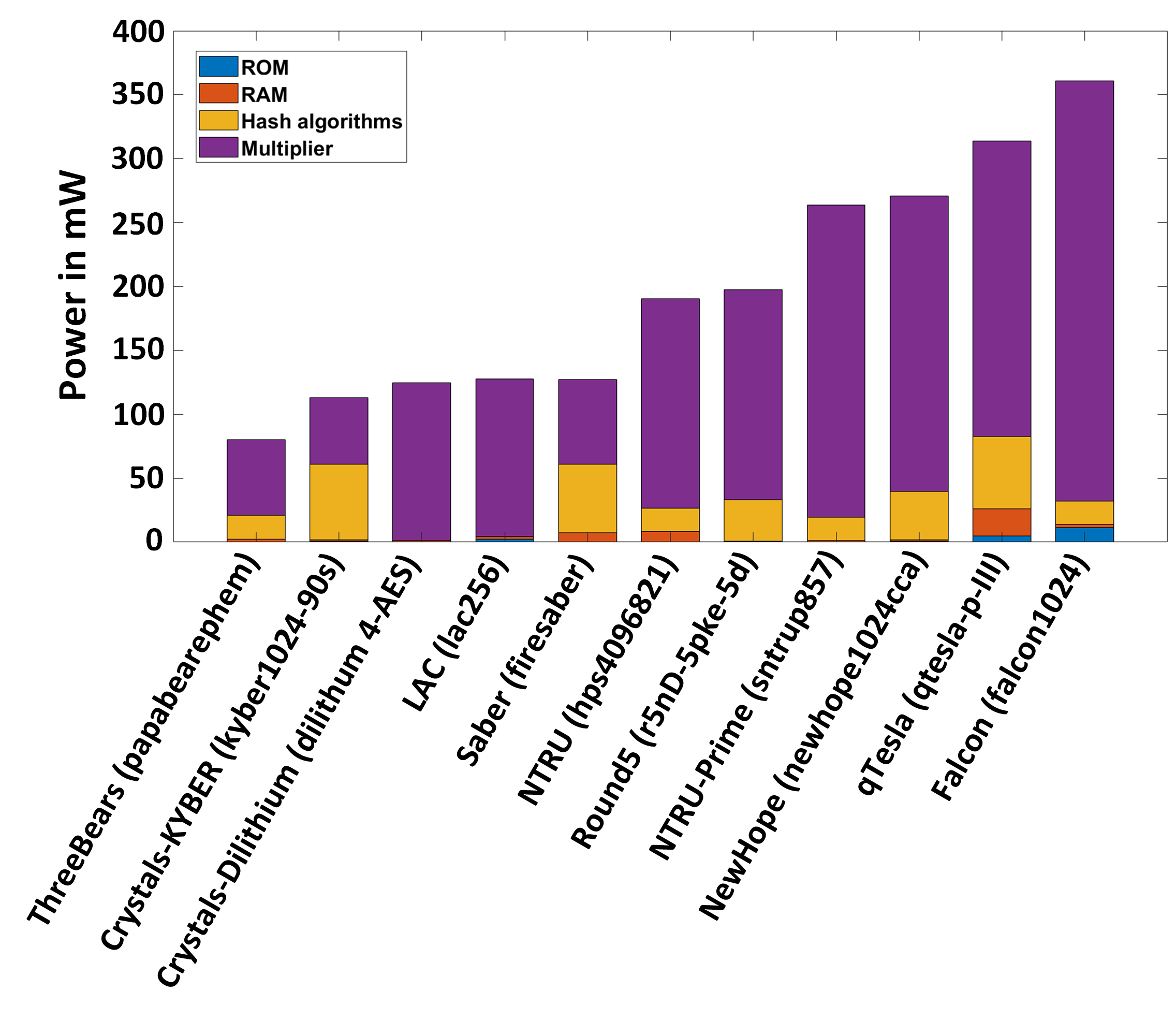}
\caption{Total power consumption of the studied NIST lattice-based PQC algorithms, ordered.}
\centering
\label{fig:figure4}
\end{figure}

The aggregate results for area and power are presented in figures \ref{fig:figure3} and \ref{fig:figure4}, respectively. We have ordered the algorithms according to their area and power requirements. Therefore, the algorithm order presented in Fig. \ref{fig:figure3} does not necessarily match the order in Fig. \ref{fig:figure4}. 

It stands out that the multiplier takes most the area resources and contributes the most to the power consumption. As we described earlier in this work, having large multipliers is a common requirement for several of the NIST PQC candidates. There are many multiplier architectures that can be used, including solutions that rely on digitized computation \cite{Rafferty2017,multipliers}. These multipliers should be adapted to the characteristics of each algorithm and to the application requirements. 

However, to make a fair comparison of the selected algorithms, we have elected to use only one multiplier in this work. Our main motivation is to provide sufficient details that can help the security community to select an appropriate algorithm according to the complexity of targeted application. The information provided in figures \ref{fig:figure3} and \ref{fig:figure4} is very useful in that regard, while also identifying which building block is the best candidate for being replaced, optimized, or even offloaded elsewhere.

Regarding area, our analysis reveals that ThreeBears is the only accelerator that has an area requirement below 0.5mm$^2$. Its area is mostly a function of the employed multiplier while the hashing core utilized requires a fraction of the resources. Memories are barely a factor. The area consumption of Crystals-KYBER and Saber lies between  0.5mm$^2$ < area < 1mm$^2$. Similarly, the area consumption of LAC, Crystals-Dilithium, and NTRU lies between 1.0mm$^2$ < area < 2.0mm$^2$. Accelerators such as  Round5, NTRU-Prime, NewHope, qTesla, and Falcon require between 2mm$^2$ and 6mm$^2$ of silicon area. In modern chip design, IP size is often measured in units of mm$^2$, while the overall chip size can be measured in tens of mm$^2$. In this sense, the algorithms aforementioned could be considered in their current form. On the other hand, FrodoKEM is too far above a \textit{reasonable size} and would be a perfect candidate for a digitized multiplier architecture. We opt not to show FrodoKEM in these charts.

As shown in Fig. \ref{fig:figure4}, the highest power consumption comes from the Falcon algorithm. Keeping the descending order, qTesla is second highest power hungry algorithm. Often, multiplier contribute high part in the total power consumption of an algorithm, but the power profile of Crystals-KYBER is completely different as it consumes \textit{112.77mW} of power. Here, the hash core is responsible for approximately 53\% of the power consumption (\textit{59.01mW} out of \textit{112.77mW}). This power figure could be reduced if the hashing function was replaced. ThreeBears is least power consumption candidate and it requires power of 79.92mW. Saber, LAC and Crystals-Dilithium accelerators consume power in the range of \textit{100mW} to \textit{150mW}. The remaining algorithms require the power between \textit{150mW} < power < \textit{400mW} and and the power profile of the power look like a exponentially increasing function -- we compare these values with other implementations in the next subsection.. Fig. \ref{fig:figure4} reveals that the power consumption is tightly linked to the consumption of the multiplier and the hashing cores. In other words, both RAMs and ROMs have little bearing on the power budget. The contribution of memories to the accelerator area is more pronounced, even if the total area is still strongly dominated by the multiplier area. Regarding frequency of operation, the bottleneck in all of our experiments is always the multiplier.

\subsection{Comparison to state-of-the-art implementations} \label{sec:performance_comparison}
After providing descriptions and evaluations of critical building blocks of NIST lattice-based PQC candidates in Section \ref{sec:Assessment_of_Reference_Implementations} and Section \ref{sec:4}, now we have to frame it with respect to state-of-the-art implementations of lattice-based algorithms in hardware. Moreover, it will help ASIC designers to select an appropriate algorithm according to the complexity of targeted application.

As described in Section \ref{sec:Intro_challenges}, the available literature on NIST lattice-based PQC candidates considers different implementation platforms \cite{Energy_Consumption_S/w,Pwr_Analysis_NTRU_Prime_S/w,H/w_study_of_sig_schemes_FPGA,Impl_Benchmark_FPGA,qTesla_FPGA,Architecture_of_NewHope_on_FPGA,Seven_lattice_Algos_FPGA_ASIC,PASS_FPGA_ASIC,SAPPHIRE,VPQC}. The implementations described in \cite{Energy_Consumption_S/w,Pwr_Analysis_NTRU_Prime_S/w} are software based and provide no good basis for comparison. Similarly, \cite{H/w_study_of_sig_schemes_FPGA,Impl_Benchmark_FPGA, qTesla_FPGA, Architecture_of_NewHope_on_FPGA} performed FPGA implementations in an Artix-7 device. ASIC implementations are described in \cite{Seven_lattice_Algos_FPGA_ASIC,PASS_FPGA_ASIC,SAPPHIRE,VPQC} where different technologies (i.e., 65nm, 40nm, and 28nm) have been used to draw area and power figures.

Moreover, the results reported in \cite{Seven_lattice_Algos_FPGA_ASIC, PASS_FPGA_ASIC} use the same 65nm node as we do in our experiments. The authors have provided results for Crystals-Kyber, NewHope, FrodoKEM, NTRU, Saber, Crystals-Dilithium, and qTesla. Except for FrodoKEM, we provide comparisons to these accelerators in Table \ref{tab:table10}. 


\begingroup
\setlength{\tabcolsep}{1.5pt} 
\renewcommand{\arraystretch}{1.5} 
\fontsize{8}{10}\selectfont
\begin{longtable}{|p{2.3cm}|p{2.7cm}|p{0.6cm}|p{1.3cm}|p{1.4cm}|p{2.1cm}|p{2.3cm}|}
\caption{Implementation details of NIST lattice-based PQC candidates on 65nm ASIC technology} 
\label{tab:table10} \\ \hline
\parbox{1.5cm}{\textbf{Ref.}} & 
\parbox{2.7cm}{\textbf{Implemented\\accelerator}} & 
\parbox{1cm}{\textbf{SL\textsubscript{i}}} & 
\parbox{1.3cm}{\textbf{Clk.\\Period\\(in \textit{ns})}} & 
\parbox{1.4cm}{\textbf{Freq.\\(in \textit{MHz})}} &
\parbox{2.4cm}{\textbf{Total area\\(in \textit{$\mu$m$^2$})}} &
\parbox{2.0cm}{\textbf{Total power\\(in \textit{mW})}} \\ \hline
\multirow{6}{*}{\cite{Seven_lattice_Algos_FPGA_ASIC}} & Crystals-KYBER & 1 & 5.0 & 200 & 3378515 & 39.21 \\ \cline{2-7}
{} & NewHope            & 1 & 5.9 & 168.6  & 3208999 & 38.02 \\ \cline{2-7}
{} & NTRU               & 1 & 5.8 & 169.5  & 1246869 & 14.30  \\ \cline{2-7}
{} & Saber              & 3 & 7.2 & 137.75 & 4774529 & 54.49 \\ \cline{2-7}
{} & Crystals-Dilithium & 1 & 6.3 & 157.7  & 4774529 & 51.24 \\ \hline 
\multirow{2}{*}{\cite{PASS_FPGA_ASIC}} & qTesla              & 1 & 5.0 & 200 & 3450765 & 16.08 \\ \cline{2-7}
{}                                     & Crystals-Dilithium  & 1 & 5.0 & 200 & 3677434 & 11.31 \\ \hline

\multirow{7}{*}{\parbox{2.2cm}{\textbf{This work}}}             & Crystals-KYBER     & 5 & 5.0 & 200 & 596300 (-82\%) & 47.39 (+21\%) \\ \cline{2-7}
{}                                     & NewHope            & 5 & 5.93 & 168.6 & 2384120 (-26\%) & 98.54 (+159\%) \\ \cline{2-7}
{}                                     & NTRU               & 5 & 5.89 & 169.5 & 1642730 (+32\%) & 78.85 (+451\%) \\ \cline{2-7}
{}                                     & Saber              & 5 & 7.3 & 137.75 & 834200 (-82\%) & 42.14 (-23\%)\\ \cline{2-7}
{}                                     & Crystals-Dilithium & 4 & 6.34 & 157.7 & 1153800 (-76\%) & 50.94 (-1\%) \\ \cline{2-7}
{}                                     & qTesla             & 3 & 5.0 & 200 & 3348300 (-3\%)& 156.44 (+873\%) \\ \cline{2-7}
{}                                     & Crystals-Dilithium & 4 & 5 & 200 & 1165100 (-68\%) & 69.41 (+513\%) \\ \hline

\multirow{2}{*}{\parbox{2.2cm}{\textbf{This work} (digitized multipliers)}}             &  qTesla (8 digits)             & 3 & 5.0 & 200 & 2187300 (-37\%)& 137.29 (+754\%) \\ \cline{2-7} {}               
 & qTesla (16 digits)             & 3 & 5.0 & 200 & 2088300 (-39\%)& 131.79 (+719\%) \\ \hline               
\end{longtable}
\endgroup

It is noteworthy that implementations of NTRU in \cite{Seven_lattice_Algos_FPGA_ASIC} require 32\% more area than our estimation while for remainig candidates our estimated area values are much lower than the implementations described in \cite{Seven_lattice_Algos_FPGA_ASIC, PASS_FPGA_ASIC}. This can be attributed mostly to the fact that we focus on building blocks and disregard the `glue logic' between them.

As shown in last column of Table \ref{tab:table10}, the power values achieved in this work are much higher than the counterparts reported in \cite{Seven_lattice_Algos_FPGA_ASIC, PASS_FPGA_ASIC}. This is due to make use of HLS for generating RTL codes in \cite{Seven_lattice_Algos_FPGA_ASIC, PASS_FPGA_ASIC}, which implies the C/C++ routines written for multipliers are operated in a loop fashion where the input and output parameters are \textit{uint16\_t/uint32\_t/uint64\_t}. Such solution is no different than a digitized/segmented multiplier, which should decrease area and power at the cost of execution time/latency. Thus, we have also provided area and power values for a segmented SBM multiplier with segment sizes of 8 and 16 digits for qTesla (16384$\times$16384). The reduced area and power values relative to the segmented versions of SBM multiplier are shown in the bottom of Table \ref{tab:table10}. The power requirements can further be reduced by increasing the number of digits/segments in the multiplier. Finally, the power requirements for the other remaining algorithms can also be reduced by utilizing segmented multipliers in their datapaths as we did for qTesla in this work.

As well as the use of larger multipliers/less segmented, the use of higher security levels in our work is also a (small) factor that results in higher hardware resources and power consumption, as shown in column three of Table \ref{tab:table10}.



\subsection{Limitations of this work } \label{sec:limitations}
Despite the fact that this paper has assessed lattice-based NIST PQC candidates based on the strict rules defined in Section \ref{sec:InEx_principles}, there are still certain limitations and caveats:
\begin{itemize}
    \item Table \ref{tab:table4} lists many of the identified arithmetic operators, but our analysis focuses on multiplication as we consider it to be more challenging than other operations. This is not enough to generate a functional crypto accelerator, but it ought be enough to capture the characteristics of it.
    There are various other operators, including transformations from one domain to another, that are required for a complete implementation of a lattice-based crypto accelerator.
    \item For each particular algorithm, there is a number of reference models that target different security levels. For each studied algorithm, we consider only the reference model with the highest security level. This approach might be an overkill for several applications that would otherwise be satisfied with an AES-128 equivalent level of security.
    \item Results for area and power are collected after logic synthesis. However, an accelerator still has to go through physical synthesis, where many additional cells are added and routing resources have to be accounted for. 
    \item We make assumptions based on reference implementations that were submitted to NIST for standardization. As the name implies, these are \textit{reference} implementations and may not be optimized on purpose for the sake of readability. Yet, we believe our approach to extract building blocks remains valid despite this limitation. 
\end{itemize}

\section{Conclusion} \label{sec:conclusion}
In this paper, we have evaluated how lattice-based algorithms participating in the NIST PQC standardization process would perform as ASIC hardware accelerators. Even if some candidates had been previously evaluated \cite{H/w_study_of_sig_schemes_FPGA,Impl_Benchmark_FPGA,Seven_lattice_Algos_FPGA_ASIC, qTesla_FPGA}, there remained many avenues of research to explore towards a fair and comprehensive comparison among algorithms. To achieve this, we have evaluated the C/C++ codes of reference implementations to extract the relevant information pertaining to this study. Furthermore, based on the extracted figures, we have compiled and generated 50+ memory instances for ROMs and RAMs with different parameter sets using a commercial memory compiler. We have also synthesized dozens of multiplier and hashing cores

From the onset, our goal was to provide ASIC designers with information that can guide their future implementations of lattice-based crypto cores. In this regard, this work highlights algorithms that would make for efficient accelerators and which building blocks could be targeted for further improvement. Nevertheless, there are many aspects that remain to be studied. In particular, in future works, we aim to investigate the multiplier design space even further and to implement unified architectures that can share resources more efficiently.

\section{Acknowledgment} \label{sec:aknowledge}
This work was supported by the EC through the European Social Fund in the context of the project ``ICT programme''.

\bibliographystyle{IEEEtran}
\bibliography{iacrdoc}

\begin{appendices}
\section{Area, power, and frequency trends for different multipliers} \label{appendix:A}

To provide representative trends, we have developed codes in Verilog for several of the multipliers employed by the reference implementations (see Section \ref{sec:Arithmetic_sizes}). Next, we performed logic synthesis of the multipliers  using a range of operand sizes. The tool utilized for this experiment is Cadence Genus. The standard cell library is a commercial one and the process node is 65nm.

\subsection{SBM, 2-Way KM, 3-Way  and 4-Way TCM multipliers} \label{appendix:A.1}

\begin{figure}[ht]
\centering \footnotesize
\includegraphics[width=4.8in,height=2.9in]{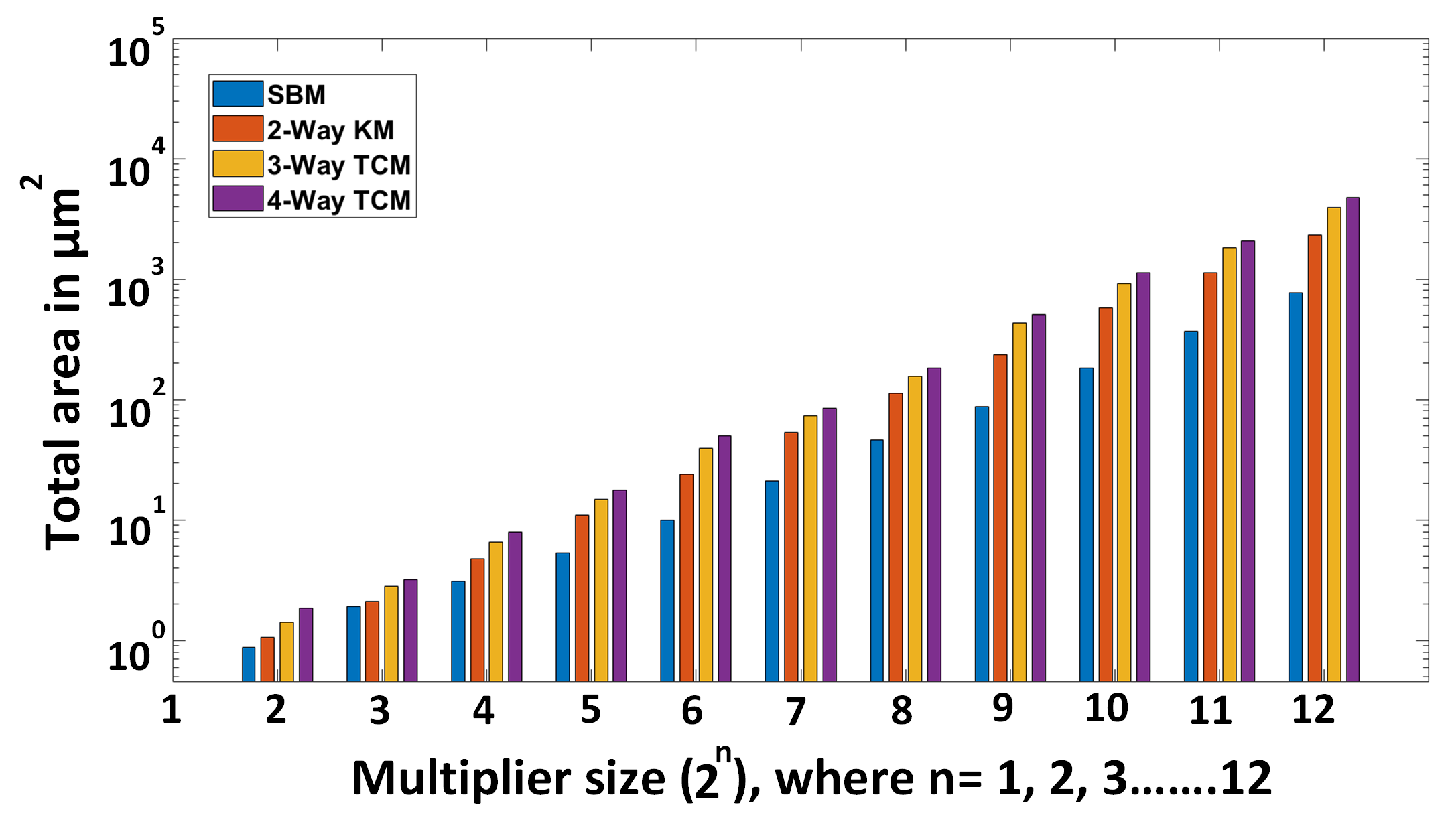}
\caption{Required area (in \textit{$\mu$m$^2$}) for different multipliers and sizes}
\centering
\label{fig:figure5}
\end{figure}

\begin{figure}[ht]
\centering \footnotesize
\includegraphics[width=4.8in,height=2.9in]{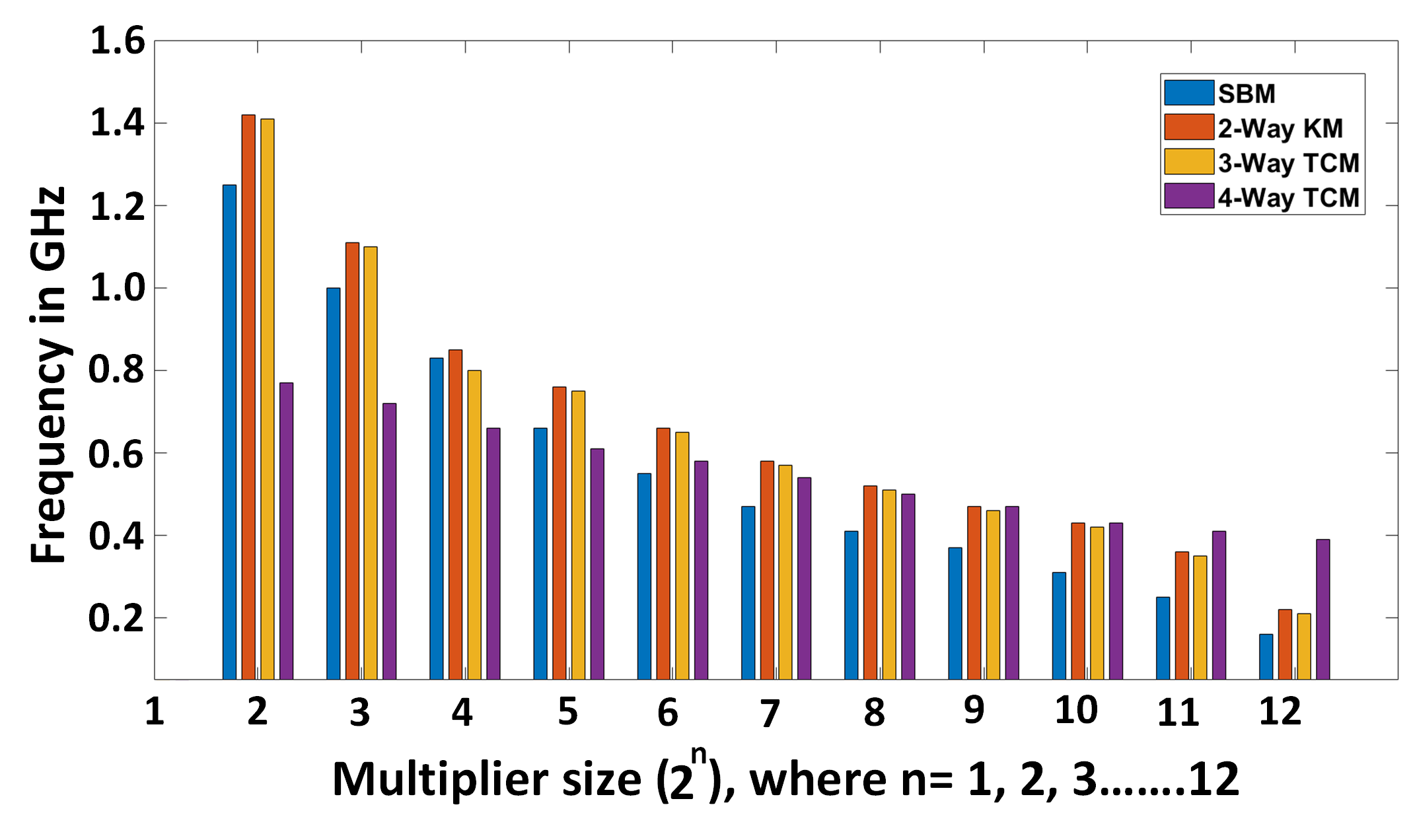}
\caption{Achieved clock frequency (in \textit{GHz}) for different multipliers and sizes}
\centering
\label{fig:figure6}
\end{figure}

\begin{figure}[ht]
\centering \footnotesize
\includegraphics[width=4.8in,height=2.9in]{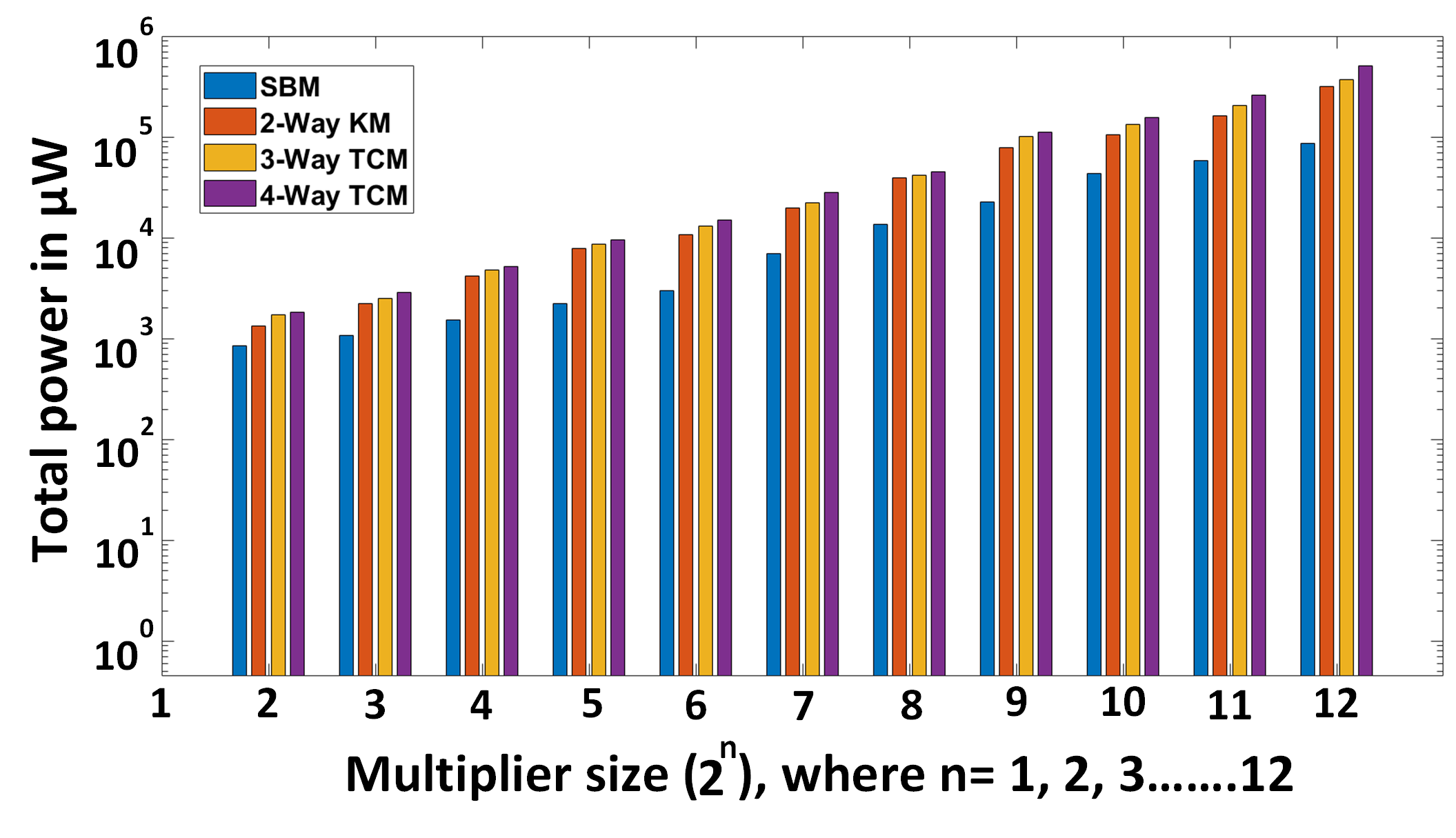}
\caption{Total power consumption (in \textit{$\mu$W}) for different multipliers and  sizes}
\centering
\label{fig:figure7}
\end{figure}

We have developed our own codes in RTL form for SBM, 2-Way KM, 3-Way TCM, and 4-Way TCM multipliers, which all take two input operands and produce one single output. The inputs are identical in size and are given in the form \textit{$2^n$}, where \textit{n} is an integer value in the range 1 to 12. In these multipliers, the output is considered without any reduction operation (i.e., the output is \textit{2$\times(2^n) - 1$} bits in length). To evaluate the performance of SBM, 2-Way KM, 3-Way TCM and 4-Way TCM multipliers, we show trends with respect to input operand sizes in figures \ref{fig:figure5}, \ref{fig:figure6}, and \ref{fig:figure7}. 

 As shown in Fig \ref{fig:figure5}, the SBM multiplier utilizes lower hardware resources in terms of area as compared to its counterparts. The 2-Way KM multiplier utilizes lower hardware resources than 3-Way and 4-Way TCM multipliers, as expected. When comparing 3-Way TCM to 4-Way TCM multipliers, the former utilizes lower hardware area than the latter. Still regarding area increase trends, it can be seen that the SBM, 2-Way KM, 3-Way TCM and 4-Way TCM multipliers show an exponential increase in size with the operand size. This is a strong argument in favor of digitized computation.
 
 For different multipliers with diverse operand lengths, the maximum achieved clock frequency is shown in Fig \ref{fig:figure6}. These values where obtained by performing synthesis runs repeatedly until the performance could no longer be improved (i.e., the constraints were pushed iteratively until they were no longer met). Based on the implementation results, it becomes clear that the 2-Way KM multiplier could be used as an efficient replacement for 3-Way TCM multiplier as it achieves comparable clock frequency while utilizing less area. For large input sizes, all multipliers outperform SBM since they benefit from breaking down the problem in chunks. Conversely, the 4-Way TCM presents the highest frequency in the comparison shown in Fig \ref{fig:figure6}. 
 
Regarding power consumption, as shown in Fig \ref{fig:figure7}, the SBM multiplier outperforms all other multipliers -- this is an important characteristic that influenced our choice of SBM as the multiplier for the PQC algorithms studied in this paper. When considering the power consumption of 2-Way KM, 3-Way TCM, and 4-Way TCM multipliers, 4-Way TCM consumes more power than 2-Way KM and 3-Way TCM multipliers. There is an increase in power consumption as the length of inputs to the multipliers are increased, as well as an increase with the number of 'ways' each multiplier uses.

\subsection{NTT multiplier} \label{appendix:A.2} For the analysis of NTT multipliers, we have adapted the pipeline architecture described in \cite{Roy2019}. Contrary to the previously discussed multipliers, here it makes little sense to decouple the pipelining from the core idea of the algorithm. In fact, it would be fair to refer to \cite{Roy2019} as a multiplier architecture instead of a simple multiplier implementation. For this reason, our NTT multiplier analysis is presented separately. We emphasize that it is a complex architecture, for which the authors carefully selected many parameters (e.g., polynomial size, residual size, number of parallel units, reduction tables, etc.) when targeting an FPGA platform. 

\begin{figure}[ht]
\centering \footnotesize
\includegraphics[width=3.0in,height=3.0in]{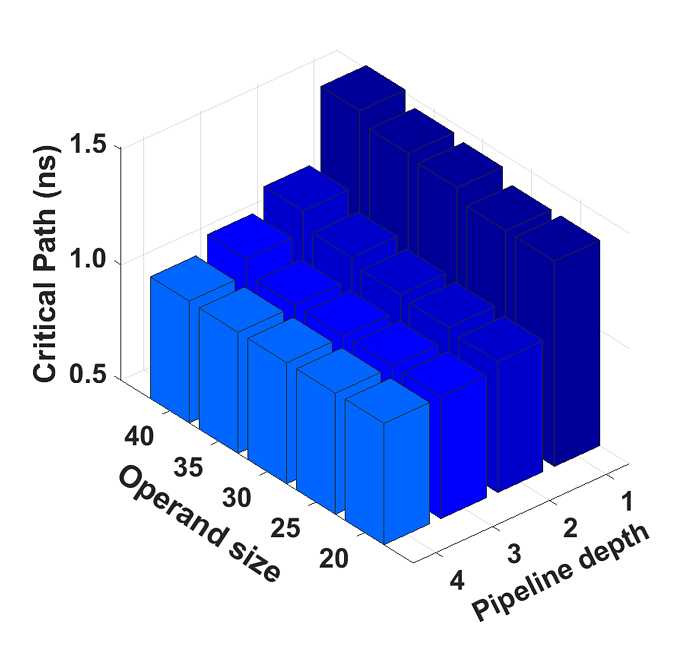}
\caption{Analysis of NTT multiplier (obtained from \cite{Roy2019} and slightly modified for ASIC)}
\centering
\label{fig:figure8}
\end{figure}

After our analysis, and in line with the results of \cite{Roy2019}, we have found the critical path of NTT multiplier to be dependent on a 30-bit integer multiplier and the reduction operation that follows it (i.e., the operation that brings the 60-bit result back to 30 bits). We have considered multiple pipeline depths (1, 2, 3 and 4) and multiple operand sizes for its integer multiplier (20, 25, 35, and 40) in our analysis \cite{harvey_2013}. In Fig \ref{fig:figure8}, we show how the critical path changes depending on the pipeline depth of the integer multiplier as well as the width of operands. From our results, it appears that the clock frequency saturates when 3 stages are used for pipelining, whereas in \cite{Roy2019} the implementation makes use of 4 stages. We believe this difference comes from the FPGA DSP unit that implements the integer multiplier in \cite{Roy2019}. In our analysis, this unit has been replaced by a ChipWare component. For the sake of providing a basis for comparison to other multipliers, we have evaluated the area required for the 30-bit/4-stage version of the NTT code, which comes to be 209740 \textit{$\mu$m$^2$} (approximately 52k cells). However, this result includes many blackboxes for the memory hierarchy that is required to implement this NTT architecture. These values should not be directly compared to other multipliers since they are known to be underestimated. 

\subsection{Summary of the multipliers trend} \label{appendix:A.3}
Due to bit serial structure, the implemented SBM multiplier requires \textit{m-1} clock cycles for \textit{m} bit operands length. The implemented 2-Way KM, 3-Way TCM, and 4-Way TCM multiplier require \textit{m/2-1}, \textit{m/3-1}, and \textit{m/4-1} clock cycles. This reduction comes with a cost in resources, naturally. Implementation of NTT multiplier requires \textit{2m+2log\textsubscript{2}$^n$} clock cycles, where \textit{n} determines the number of NTT points. 

The performance of the evaluated multipliers could be improved significantly by using different optimization techniques, most notably pipelining and digitizing. However, in these plots (Fig. \ref{fig:figure5}--\ref{fig:figure7}), we have not provided optimized results either for pipelining or digitizing, as we are interested in showing trends that capture the core idea of each algorithm.

Consequently, there is always a trade-off when selecting an appropriate multiplier and evaluating its area, frequency, and power characteristics. In this work, the SBM multiplier was deemed more appropriate due, its flexibility, its lower area footprint, and its lower power consumption (with respect to the other multipliers we considered).

Finally, regarding NTT-based multipliers, while it appears tremendously advantageous, generating a single NTT multiplier architecture that would be a good fit for all NIST candidates is not feasible. The parameter space requires careful algorithm-specific exploration.
\end{appendices}
\end{document}